\documentclass[]{article}
\usepackage{lmodern}
\usepackage[table,dvipsnames]{xcolor}
\usepackage{amssymb,amsmath}
\usepackage{ifxetex,ifluatex}
\usepackage{fixltx2e} 
\ifnum 0\ifxetex 1\fi\ifluatex 1\fi=0 
  \usepackage[T1]{fontenc}
  \usepackage[utf8]{inputenc}
\else 
  \ifxetex
    \usepackage{mathspec}
  \else
    \usepackage{fontspec}
  \fi
  \defaultfontfeatures{Ligatures=TeX,Scale=MatchLowercase}
\fi
\IfFileExists{upquote.sty}{\usepackage{upquote}}{}
\IfFileExists{microtype.sty}{%
\usepackage{microtype}
\UseMicrotypeSet[protrusion]{basicmath} 
}{}
\usepackage[margin=1in]{geometry}
\usepackage{hyperref}
\PassOptionsToPackage{usenames,dvipsnames}{color} 
\hypersetup{unicode=true,
            pdftitle={Indirect inference through prediction},
            colorlinks=true,
            linkcolor=Maroon,
            citecolor=Blue,
            urlcolor=blue,
            breaklinks=true}
\usepackage{longtable,booktabs}
\usepackage[singlelinecheck=off]{caption}
\usepackage{graphicx,grffile}
\makeatletter
\def\maxwidth{\ifdim\Gin@nat@width>\linewidth\linewidth\else\Gin@nat@width\fi}
\def\maxheight{\ifdim\Gin@nat@height>\textheight\textheight\else\Gin@nat@height\fi}
\makeatother
\setkeys{Gin}{width=\maxwidth,height=\maxheight,keepaspectratio}
\IfFileExists{parskip.sty}{%
\usepackage{parskip}
}{
\setlength{\parindent}{0pt}
\setlength{\parskip}{6pt plus 2pt minus 1pt}
}
\providecommand{\tightlist}{%
  \setlength{\itemsep}{0pt}\setlength{\parskip}{0pt}}
\ifx\paragraph\undefined\else
\let\oldparagraph\paragraph
\renewcommand{\paragraph}[1]{\oldparagraph{#1}\mbox{}}
\fi
\ifx\subparagraph\undefined\else
\let\oldsubparagraph\subparagraph
\renewcommand{\subparagraph}[1]{\oldsubparagraph{#1}\mbox{}}
\fi

\usepackage[section]{placeins}
\let\Oldsection\section
\renewcommand{\section}{\FloatBarrier\Oldsection}

\let\Oldsubsection\subsection
\renewcommand{\subsection}{\FloatBarrier\Oldsubsection}

\let\Oldsubsubsection\subsubsection
\renewcommand{\subsubsection}{\FloatBarrier\Oldsubsubsection}

\let\rmarkdownfootnote\footnote%
\def\footnote{\protect\rmarkdownfootnote}

\usepackage{titling}

  \title{Indirect inference through prediction}
  \pretitle{\vspace{\droptitle}\centering\huge}
  \posttitle{\par}
  \author{Ernesto Carrella \\ Richard M. Bailey \\ Jens Koed Madsen \\
  \scriptsize{School of Geography and the Environment, University of Oxford, South Parks Road, Oxford, OX1 3QY, UK.}}
  \preauthor{\centering\large\emph}
  \postauthor{\par}
  \predate{\centering\large\emph}
  \postdate{\par}
  \date{2018-07-04}

\usepackage{booktabs}
\usepackage{longtable}
\usepackage{array}
\usepackage{multirow}

\usepackage{wrapfig}
\usepackage{float}
\usepackage{colortbl}
\usepackage{pdflscape}
\usepackage{tabu}
\usepackage{threeparttable}
\usepackage{threeparttablex}
\usepackage[normalem]{ulem}
\usepackage{makecell}

\usepackage{algorithmic}
\usepackage{algorithm}

\usepackage{amsthm}

\theoremstyle{definition}

\theoremstyle{definition}

\theoremstyle{definition}

\theoremstyle{remark}

\begin{document}
\maketitle
\begin{abstract}
By recasting indirect inference estimation as a prediction rather than a
minimization and by using regularized regressions, we can bypass the
three major problems of estimation: selecting the summary statistics,
defining the distance function and minimizing it numerically. By
substituting regression with classification we can extend this approach
to model selection as well. We present three examples: a statistical
fit, the parametrization of a simple real business cycle model and
heuristics selection in a fishery agent-based model. The outcome is a
method that automatically chooses summary statistics, weighs them and
use them to parametrize models without running any direct minimization.
\end{abstract}

{
\hypersetup{linkcolor=black}
\setcounter{tocdepth}{2}
}
\hypertarget{jel-classification}{%
\section*{JEL Classification}\label{jel-classification}}
\addcontentsline{toc}{section}{JEL Classification}

C15; C63; C3

\hypertarget{keywords}{%
\section*{Keywords}\label{keywords}}
\addcontentsline{toc}{section}{Keywords}

Agent-based models; Indirect inference; Estimation; Calibration;
Simulated minimum distance;

\hypertarget{acknowledgments}{%
\section*{Acknowledgments}\label{acknowledgments}}
\addcontentsline{toc}{section}{Acknowledgments}

This research is funded in part by the Oxford Martin School, the David
and Lucile Packard Foundation, the Gordon and Betty Moore Foundation,
the Walton Family Foundation, and Ocean Conservancy.

\hypertarget{introduction}{%
\section{Introduction}\label{introduction}}

We want to tune the parameters of a simulation model to match data. When
the likelihood is intractable and the data high-dimensional, the common
approach is the simulated minimum distance (Grazzini and Richiardi
\protect\hyperlink{ref-Grazzini2015}{2015}). This involves summarising
real and simulated data into a set of summary statistics and then tuning
the model parameters to minimize their distance.

Minimization involves three tasks. First, choose the summary statistics.
Second, weight their distance. Third, perform the numerical minimization
keeping in mind that simulation may be computationally expensive.

Conveniently, ``regression adjustment methods'' from the approximate
Bayesian computation(ABC) literature (Blum et al.
\protect\hyperlink{ref-Blum2013}{2013}) can be adapted to side-step all
3 tasks. Moreover we can use these techniques outside of the ABC
super-structure.

We substitute the minimization with a regression. First, run the model
``many'' times and observe pairs of model parameters and generated
summary statistics. Then, regress the parameters as the dependent
variable against all the simulated summary statistics. Finally, plug in
the real summary statistics in the regression to obtain the tuned
parameters. Model selection can be similarly achieved by training a
classifier instead.

While the regressions can in principle be non-linear and/or
non-parametric, we limit ourselves here to regularized linear
regressions (the regularization is what selects the summary statistics).
In essence, we can feed the regression a long list of summary statistics
we think relevant and have the regularization select only what matters
for parametrization.

We summarize the agent-based estimation in general and regression
adjustment methods in particular in section \ref{litreview}. We describe
how to parametrize models in section \ref{method}. We then provide 3
examples: a statistical line fit in section \ref{linerexample}, a real
business cycle macroeconomic model in section \ref{rbcexample} and a
fishery agent-based model in section \ref{abmexample}. We conclude by
pointing out weaknesses and possible extensions in section
\ref{concludes}.

\hypertarget{litreview}{%
\section{Literature review}\label{litreview}}

The approach of condensing data and simulations into a set of summary
statistics in order to match them is covered in detail in two review
papers: Hartig et al.
(\protect\hyperlink{ref-hartig_statistical_2011}{2011}) in ecology and
Grazzini and Richiardi (\protect\hyperlink{ref-Grazzini2015}{2015}) in
economics.\\
They differ only in their definition of summary statistics: Hartig et
al. (\protect\hyperlink{ref-hartig_statistical_2011}{2011}) defines it
as any kind of observation available in both model and reality while
Grazzini and Richiardi (\protect\hyperlink{ref-Grazzini2015}{2015}),
with its focus on consistency, allows only for statistical moments or
auxiliary parameters from ergodic simulations.

Grazzini and Richiardi (\protect\hyperlink{ref-Grazzini2015}{2015}) also
introduces the catch-all term ``simulated minimum distance'' for a set
of related estimation techniques, including indirect
inference(Gourieroux, Monfort, and Renault
\protect\hyperlink{ref-Gourieroux1993}{1993}) and the simulated method
of moments(McFadden \protect\hyperlink{ref-McFadden1989}{1989}). Smith
(\protect\hyperlink{ref-Smith1993}{1993}) first introduced indirect
inference to fit a 6-parameter real business cycle macroeconomic
model.\\
In agent-based models, indirect inference was first advocated in
Richiardi et al. (\protect\hyperlink{ref-Richiardi2006}{2006}; see also
Shalizi \protect\hyperlink{ref-Shalizi2017}{2017}) and first implemented
in Zhao (\protect\hyperlink{ref-Zhao2010}{2010}) (who also proved mild
conditions for consistency).

Common to all is the need to accomplish three tasks: selecting the
summary statistics, defining the distance function and choosing how to
minimize it.

Defining the distance function may at first appear the simplest of these
tasks. It is standard to use the Euclidean distance between summary
statistics weighed by the inverse of their variance or covariance
matrix. However, while asymptotically efficient, this was shown to
underperform in Smith (\protect\hyperlink{ref-Smith1993}{1993}) and
Altonji and Segal (\protect\hyperlink{ref-Altonji1996}{1996}). The
distance function itself is a compromise, since in all but trivial
problems we face a multi-objective optimization(see review by Marler and
Arora \protect\hyperlink{ref-Marler2004}{2004}) where we should minimize
the distance to each summary statistic independently. In theory, we
could solve this by searching for an optimal Pareto front (see Badham et
al. \protect\hyperlink{ref-Badham2017}{2017} for an example of dominance
analysis applied to agent-based models). In practice, however, even a
small number of summary statistics make this approach infeasible and
unintelligible. Better then to focus on a single imperfect distance
function (a weighted global criterion, in multi-objective optimization
terms) where there is at least the hope of effective minimization. A
practical solution that may nonetheless obscure tradeoffs between
summary statistic distances.

We can minimize distance function with off-the-shelf methods such as
genetic algorithms (Calver and Hutzler
\protect\hyperlink{ref-Calver2006}{2006}; Heppenstall, Evans, and Birkin
\protect\hyperlink{ref-Heppenstall2007}{2007}; Lee et al.
\protect\hyperlink{ref-Lee2015}{2015}) or simulated annealing (Le et al.
\protect\hyperlink{ref-Le2016}{2016}; Zhao
\protect\hyperlink{ref-Zhao2010}{2010}). More recently two Bayesian
frameworks have been prominent: BACCO (Bayesian Analysis of Computer
Code Output) and ABC (Approximate Bayesian Computation).

BACCO involves running the model multiple times for random parameter
inputs, building a statistical meta-model (sometimes called
\emph{emulator} or \emph{surrogate}) predicting distance as a function
of parameters and then minimizing the meta-model as a short-cut to
minimizing the real distance. Kennedy and O'Hagan
(\protect\hyperlink{ref-Kennedy2001a}{2001}) introduced the method(see
O'Hagan \protect\hyperlink{ref-OHagan2006}{2006} for a review).
Ciampaglia (\protect\hyperlink{ref-Ciampaglia2013}{2013}), Parry et al.
(\protect\hyperlink{ref-Parry2013}{2013}) and Salle and Yıldızoğlu
(\protect\hyperlink{ref-Salle2014}{2014}) are recent agent-based model
examples.

A variant of this approach interleaves running simulations and training
the meta-model to sample more promising areas and achieve better minima.
The general framework is the ``optimization by model fitting''(see
chapter 9 of Sean \protect\hyperlink{ref-luke_essentials_2009}{2010} for
a review; see Michalski \protect\hyperlink{ref-Michalski2000}{2000} for
an early example). Lamperti, Roventini, and Sani
(\protect\hyperlink{ref-Lamperti2018}{2018}) provides an agent-based
model application (using gradient trees as a meta-model). Bayesian
optimization (see Shahriari et al.
\protect\hyperlink{ref-shahriari_taking_2016}{2016} for a review) is the
equivalent approach using Gaussian processes and Bailey et al.
(\protect\hyperlink{ref-Bailey2018}{2018}) first applied it to
agent-based models.

As in BACCO, ABC methods also start by sampling the parameter space at
random but they only retain simulations whose distance is below a
pre-specified threshold in order to build a posterior distribution of
acceptable parameters(see Beaumont
\protect\hyperlink{ref-Beaumont2010}{2010} for a review). Drovandi,
Pettitt, and Faddy (\protect\hyperlink{ref-Drovandi2011}{2011}),
Grazzini, Richiardi, and Tsionas
(\protect\hyperlink{ref-Grazzini2017}{2017}) and Zhang et al.
(\protect\hyperlink{ref-Zhang2017}{2017}) apply the method to
agent-based models. Zhang et al.
(\protect\hyperlink{ref-Zhang2017}{2017}) notes that ABC matches the
``pattern oriented modelling'' framework of Grimm et al.
(\protect\hyperlink{ref-Grimm2005}{2005}).

All these algorithms however require the user to select the summary
statistics first. This is always an \emph{ad-hoc} procedure. Beaumont
(\protect\hyperlink{ref-Beaumont2010}{2010}) describes it as ``rather
arbitrary'' and notes that its effects ``{[}have{]} not been
systematically studied''. The more summary statistics we use, the more
informative the simulated distance is (see Bruins et al.
\protect\hyperlink{ref-Bruins2018}{2018} which compares four auxiliary
models of increasing complexity). However the more summary statistics we
use the harder the minimization (and the more important the weighting of
individual distances) becomes.

Because of its use of kernel smoothing, ABC methods are particularly
vulnerable to the ``curse of dimensionality'' with respect to the number
of summary statistics. As a consequence the ABC literature has developed
many ``dimension reduction'' techniques (see Blum et al.
\protect\hyperlink{ref-Blum2013}{2013} for a review). Of particular
interest here is the ``regression adjustment'' literature (Beaumont,
Zhang, and Balding \protect\hyperlink{ref-Beaumont2002}{2002}; Blum and
Francois \protect\hyperlink{ref-Blum2010}{2010}; Nott et al.
\protect\hyperlink{ref-Nott2014}{2011}) which advocates various ways to
build regressions to map parameters from the summary statistics they
generate. Our paper simplifies and applies this approach to agent-based
models.

Model selection in agent-based models has received less attention. As in
statistical learning one can select models on the basis of their
out-of-sample predictive power (see Chapter 8 of Hastie, Tibshirani, and
Friedman \protect\hyperlink{ref-friedman_elements_2001}{2009}). The
challenge is then to develop a single measure to compare models of
different complexity making multi-variate predictions. This has been
fundamentally solved in Barde (\protect\hyperlink{ref-Barde2017}{2017})
by using Markovian Information Criteria.\\
Our approach in section \ref{patternrecognition} differs from standard
statistical model selection by treating it as a pattern recognition
problem (that is, a classification), rather than making the choice based
on ranking of out-of-sample prediction performance.

Fagiolo, Moneta, and Windrum (\protect\hyperlink{ref-Fagiolo2007}{2007})
were the first to tackle the growing variety of calibration methods in
agent-based models. In the ensuing years, this variety has grown
enormously. In the next section we add to this problem by proposing yet
another estimation technique.

\hypertarget{method}{%
\section{Method}\label{method}}

\hypertarget{estimation}{%
\subsection{Estimation}\label{estimation}}

Take a simulation model depending on \(K\) parameters
\(\theta = \left[ \theta_1,\theta_2,\dots,\theta_K \right]\) to output
\(M\) simulated summary statistics
\(S(\theta) = \left[ S_1(\theta), S_2(\theta),\dots,S_M(\theta) \right]\).
Given real-world observed summary statistics \(S^*\) we want to find the
parameter vector \(\theta^*\) that generated them.\\
Our method proceeds as follows:

\begin{enumerate}
\def\labelenumi{\arabic{enumi}.}
\tightlist
\item
  Repeatedly run the model each time supplying it a random vector
  \(\hat \theta\)
\item
  Collect for each simulation its output as summary statistics
  \(S(\hat \theta)\)
\item
  Run \(K\) separate regularized regressions, \(r_i\), one for each
  parameter. The parameter is the dependent variable and all the
  candidate summary statistics are independent variables: 
  \begin{equation}
  \left\{\begin{matrix}
  \theta_1 = r_1(S_1,S_2,\dots,S_M)\\ 
  \theta_2 = r_2(S_1,S_2,\dots,S_M)\\ 
  \vdots \\
  \theta_n = r_n(S_1,S_2,\dots,S_M)
  \end{matrix}\right.  \label{eq:regressions}  
  \end{equation}
\item
  Plug in the ``real'' summary statistics \(S^*\) in each regression to
  predict the ``real'' parameters \(\theta^*\)
\end{enumerate}

In step 1 and 2 we use the model to generate a data-set: we repeatedly
input random parameters \(\hat \theta\) and we collect the output
\(\hat S_1,\dots,\hat S_M\) (see Table \ref{tab:sample} for an example).
In step 3 we take the data we generated and use it to train regressions
(one for each model input) where we want to predict each model input
\(\theta\) looking only at the model output \(S_1,\dots,S_M\). The
regressions map observed outputs back to the inputs that generated
them.\\
Only in step 4 we use the real summary statistics (which we have ignored
so far) to plug in the regressions we built in step 3. The prediction
the regressions make when we feed in the real summary statistics \(S^*\)
are the estimated parameters \(\theta^*\).

\begin{table}
\caption{\label{tab:sample} A sample table generated after step 1 and 2 of
the estimation procedure (assuming the simulation model depends on a
single parameter \(\theta_1\) ). Each row is a separate simulation run
where \(\hat \theta_1\) was the input of the model and
\(\hat S_1,\dots,\hat S_M\) was its output. In step 3 we use this table
to train a regression that uses as explanatory variables the output of
the model \(S_1,\dots,S_M\) and try to predict from them the model input
that generated them \(\theta_1\)}
\centering
\begin{tabular}[t]{ccccc}
\hline
\(\hat \theta_1\) & \(\hat S_1\) & \(\hat S_2\) & \(\dots\) &
\(\hat S_M\)\\
\hline
\(\vdots\) & \(\vdots\) & \(\vdots\) & \(\vdots\) &
\(\vdots\)\\
\(\vdots\) & \(\vdots\) & \(\vdots\) & \(\vdots\) &
\(\vdots\)\\
\end{tabular}
\end{table}

The key implementation detail is to use regularized regressions. We
used elastic nets (Friedman, Hastie, and Tibshirani
\protect\hyperlink{ref-Friedman2010}{2010}) because they are linear and
they combine the penalties of LASSO and ridge regressions. LASSO
penalties automatically drop summary statistics that are uninformative.
Ridge penalties lower the weight of summary statistics that are strongly
correlated to each other.\\
We can then start with a large number of summary statistics (anything we
think may matter and that the model can reproduce) and let the
regression select only the useful ones. The regressions could also be
made non-linear and non-parametric but this proved not necessary for our
examples.

Compare this to simulated minimum distance. There, the mapping between
model outputs \(S(\cdot)\) and inputs \(\theta\) is defined by the
minimization of the distance function: \begin{equation} 
\theta^* = \arg_{\theta}\min (S(\theta)-S^*) W (S(\theta)-S^*)  \label{eq:waldtype}    
\end{equation} When using a simulated minimum distance approach, both the summary
statistics \(S(\cdot)\) and their weights \(W\) need to be chosen in
advance. Then one can estimates \(\theta^*\) by explicitly minimizing
the distance. In effect the \(\arg \min\) operator maps model output
\(S(\cdot)\) to model input \(\theta\).\\
Instead, in our approach, the regularized regression not only selects
and weights summary statistics but, more importantly, also maps them
back to model input. Hence, the regression bypasses (takes the place of)
the usual minimization.

\hypertarget{patternrecognition}{%
\subsection{Model Selection}\label{patternrecognition}}

We observe summary statistics \(S^*\) from the real world and we would
like to know which candidate model \(\{m_1,\dots,m_n\}\) generated
them.\\
We can solve this problem by training a classifier against simulated
data.

\begin{enumerate}
\def\labelenumi{\arabic{enumi}.}
\tightlist
\item
  Repeatedly run each model
\item
  Collect for each simulation its generated summary statistics
  \(S(m_i)\)
\item
  Build a classifier predicting the model by looking at summary
  statistics and train it on the data-set just produced: \begin{equation}
  i \sim g(S_1,S_2,\dots,S_M) \label{eq:classifier} 
  \end{equation}
\item
  Plug in ``real'' summary statistics \(S^*\) in the classifier to
  predict which model generated it
\end{enumerate}

The classifier family \(g(\cdot)\) can also be non-linear and
non-parametric. We used a multinomial regression with elastic net
regularization(Friedman, Hastie, and Tibshirani
\protect\hyperlink{ref-Friedman2010}{2010}); its advantages is that the
regularization selects summary statistics and the output is a classic
logit formula that is easily interpretable.

Model selection cannot be done on the same data-set used for estimation.
If the parameters of a model need to be estimated, the estimation must
be performed on on either a separate training data-set or the inner loop
of a nested cross-validation(Hastie, Tibshirani, and Friedman
\protect\hyperlink{ref-friedman_elements_2001}{2009}).

\hypertarget{examples}{%
\section{Examples}\label{examples}}

\hypertarget{linerexample}{%
\subsection{Fit Lines}\label{linerexample}}

\hypertarget{regression-based-methods-fit-straight-lines-as-well-as-abc-or-simulated-minimum-distance}{%
\subsubsection{Regression-based methods fit straight lines as well as
ABC or Simulated Minimum
Distance}\label{regression-based-methods-fit-straight-lines-as-well-as-abc-or-simulated-minimum-distance}}

In this section we present a simplified one-dimensional parametrization
problem and show that regression methods perform as well as ABC and
simulated minimum distance.

We observe 10 summary statistics (in this case these are just direct observations) \(S^*=(S^*_0,\dots,S^*_9)\) and we
assume they were generated by the model \(S_i=\beta i + \epsilon\) where
\(\epsilon \sim \mathcal{N}(0,1)\). Assuming the model is correct, we
want to estimate the \(\beta^*\) parameter that generated the observed
\(S^*\). Because the model is a straight line, we could solve the
estimation by least squares. Pretend however this model is a
computational black box where the connection between it and drawing
straight lines is not apparent.

We run 1,000 simulations with \(\beta \sim U[0,2]\) (the range
  is arbitrary and could represent either a prior or a feasible
  interval), collecting the 10 simulated summary statistics
\(\hat S(\beta_i)\). We then train a linear elastic net regression of
the form: \begin{equation} 
\beta = a + \sum_{i=0}^9 b_i \hat S_i \label{eq:linelineline}  
\end{equation} Table \ref{tab:linercoeff} shows the \(b\) coefficients of the
regression. \(S_0\) contains no information about \(\beta\) (since it is
generated as \(S_0 = \epsilon\)) and the regression correctly ignores it
by dropping \(b_0\). The elastic net weights larger summary statistics
more because the effect of \(\beta\) is larger there compared to the
\(\epsilon\) noise.

\begin{table}

\caption{\label{tab:linercoeff}Coefficients generated by fitting an elastic net against a training sample of size 1000. Notice that $b_0$ has been automatically dropped since in this case $S_0$ contains no information about $\beta$}
\centering
\begin{tabular}[t]{l|r}
\hline
Term & Estimate\\
\hline
(Intercept) & 0.0297673\\
\hline
$b_1$ & 0.0010005\\
\hline
$b_2$ & 0.0065579\\
\hline
$b_3$ & 0.0056334\\
\hline
$b_4$ & 0.0109605\\
\hline
$b_5$ & 0.0134782\\
\hline
$b_6$ & 0.0190075\\
\hline
$b_7$ & 0.0203925\\
\hline
$b_8$ & 0.0238694\\
\hline
$b_9$ & 0.0281099\\
\hline
\end{tabular}
\end{table}

\begin{figure}[!htb]
\centering
\includegraphics{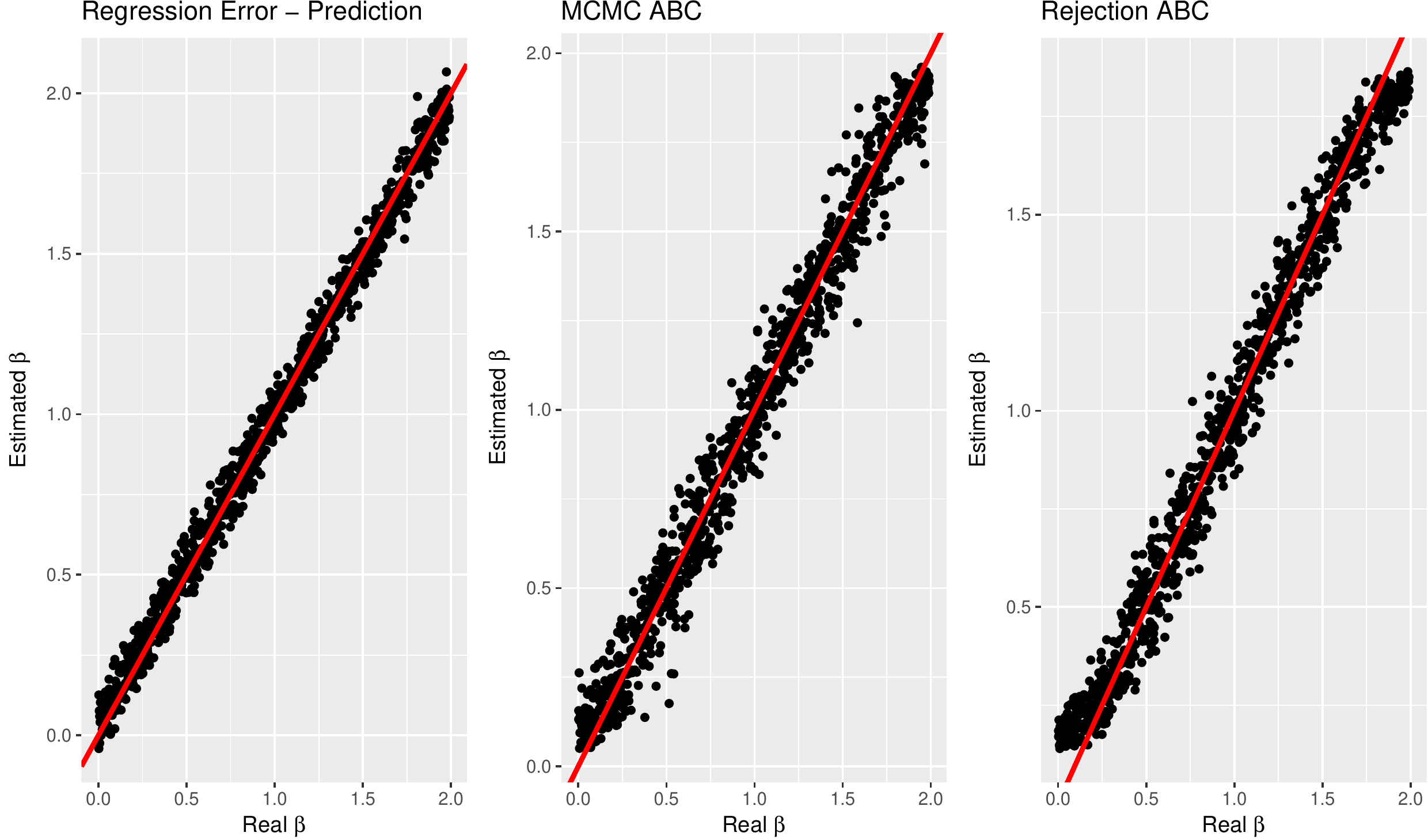}
\caption{\label{fig:linererrorplots}A side by side comparison between the
`real' and discovered \(\beta\) as generated by either the regression
method or two standard ABC algorithms: rejection sampling or MCMC. The
x-axis represents the real \(\beta\), the y-axis its estimate. Each
point represents the estimation against a different set of summary
statistics. The red line is the 45 degree line: the closer the points
are to it the better the estimation. All methods perform equally well.}
\end{figure}

We test this regression by making it predict, out-of-sample, the
\(\beta\) of 1000 more simulations looking only at their summary
statistics. Figure \ref{fig:linererrorplots} and table
\ref{tab:linererrorrate} compare the estimation quality with those
achieved by the standard rejection ABC and MCMC ABC (following Marjoram
et al. \protect\hyperlink{ref-Marjoram2003}{2003}; Wegmann, Leuenberger,
and Excoffier \protect\hyperlink{ref-Wegmann2009}{2009}; both methods
implemented in Jabot et al. \protect\hyperlink{ref-Jabot2015}{2015}).
All achieve similar error rates.

\begin{table}

\caption{\label{tab:linererrorrate}All three methods achieve on average similar estimation errors, although the regression-based method is much quicker}
\centering
\begin{tabular}[t]{l|ccr}
\hline
Estimation & Average RMSE & Runtime & Runtime(scaled)\\
\hline
Regression Method & 0.0532724 & 0.002239561 mins & 1.0000\\
\hline
Rejection ABC & 0.0823950 & 0.9586897 mins & 428.0703\\
\hline
MCMC ABC & 0.0835164 & 5.453199 mins & 2434.9408\\
\hline
\end{tabular}
\end{table}

\begin{figure}[!htb]
\centering
\includegraphics{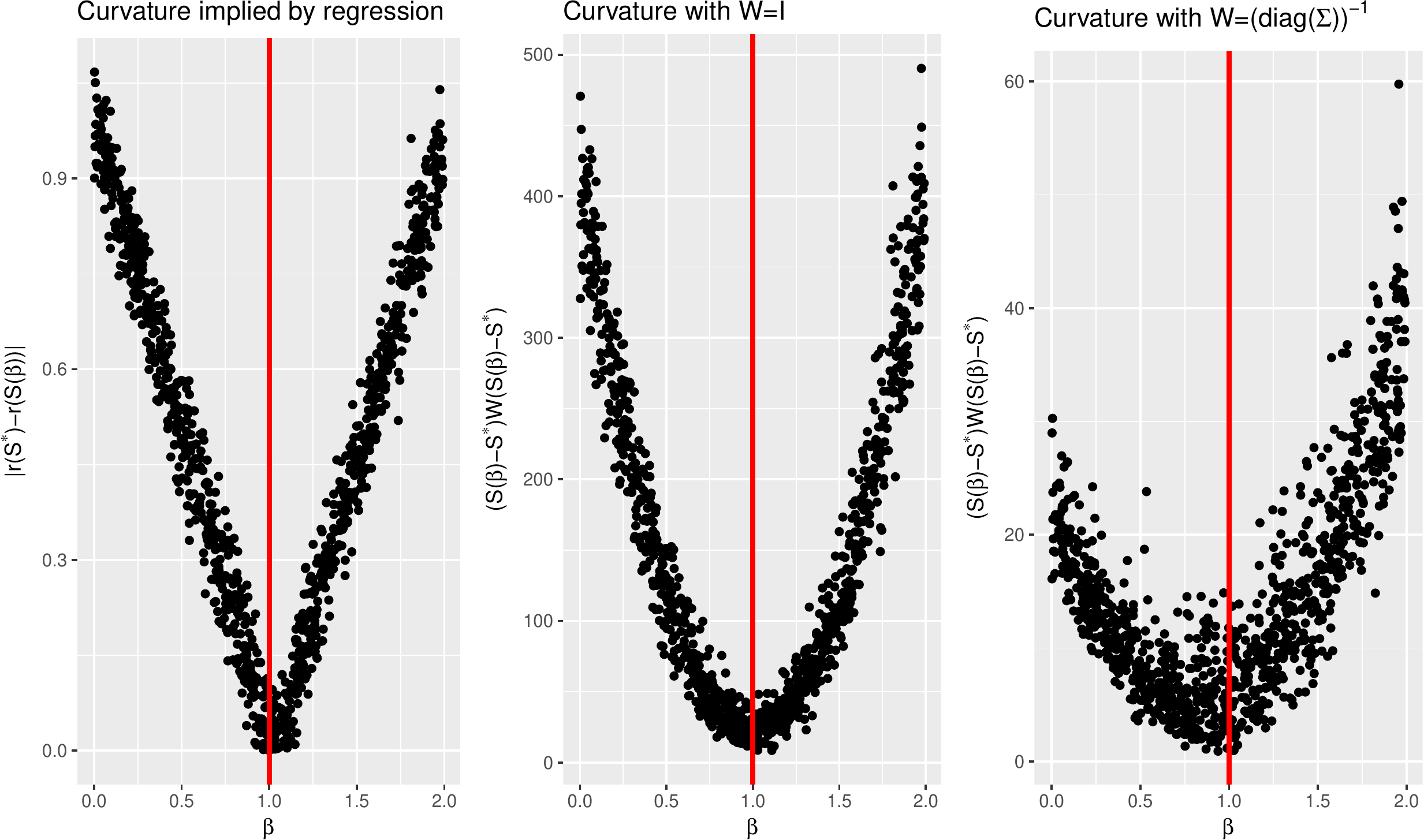}
\caption{\label{fig:linerestimationslope}Curvature of the fitness function
as implied by the regression versus weighted Euclidean distance between
summary statistics. We take the 1000 testing simulation runs, we pick
one in the middle (\(\beta \approx 1\)) as the `real' one and we compute
the summary statistics distances between all the other simulations to
it. The red line represents the real \(\beta\), each dot is the distance
to the real \(\beta\) for that simulation run and its associated
\(\beta\).}
\end{figure}

We can also compare the regression against the standard simulated
minimum distance approach. Simulated minimum distance search for the
parameters that minimize: \begin{equation} 
\beta^* = \arg_{\beta}\min (S(\beta)-S^*) W (S(\beta)-S^*)  \label{eq:waldtype}    
\end{equation} The steeper its curvature, the easier it is to minimize
numerically.\\
In figure \ref{fig:linerestimationslope} we compare the curvature
defined by \(W=I\) and \(W= \left(\text{diag}(\Sigma)\right)^{-1}\)
(where \(I\) is the identity matrix and \(\Sigma\) is the covariance
matrix of summary statistics) against the curvature implied by our
regression \(r\): \(|r(S^*)-r(S)|\) (that is, the distance between the
estimation made given the real summary statistics, \(r(S^*)\), and the
estimation made at other simulated summary statistics \(r(S)\)) . This
problem is simple enough that the usual weight matrices produce distance
functions that are easy to numerically minimize.

We showed here that regression-based methods perform as well as ABC and
simulated minimum distance in a simple one-dimensional problem.

\hypertarget{regression-based-methods-fit-broken-lines-better-than-abc-or-simulated-minimum-distance}{%
\subsubsection{Regression-based methods fit broken lines better than ABC
or simulated minimum
distance}\label{regression-based-methods-fit-broken-lines-better-than-abc-or-simulated-minimum-distance}}

In this section we present another simplified one-dimensional
parametrization but show that regression methods outperform both ABC and
simulated minimum distance because of better selection of summary
statistics.

We observe 10 summary statistics \(S=(S_0,\dots,S_9)\) but we assume
they were generated by the ``broken line'' model: \begin{equation}
S_i=\left\{\begin{matrix}
\epsilon & i < 5\\ 
\beta i + \epsilon & i\geq5
\end{matrix}\right. \label{eq:brokenline}  
\end{equation}\\
where \(\epsilon \sim \mathcal{N}(0,1)\).\\
We want to find the \(\beta\) that generated the observed summary
statistics. Notice that \(S_0,\dots,S_4\) provide no information on
\(\beta\).

As before we run the model 1000 times to train an elastic net of the
form: \begin{equation} 
\beta = a + \sum_{i=0}^9 b_i S_i \label{eq:linelineline}  
\end{equation} Table \ref{tab:brokenlinerestimation} shows the coefficients found.
The regression correctly drops \(b_0,\dots,b_4\) and weighs the
remaining summary statistics more the higher they are.

\begin{figure}[!htb]
\centering
\includegraphics{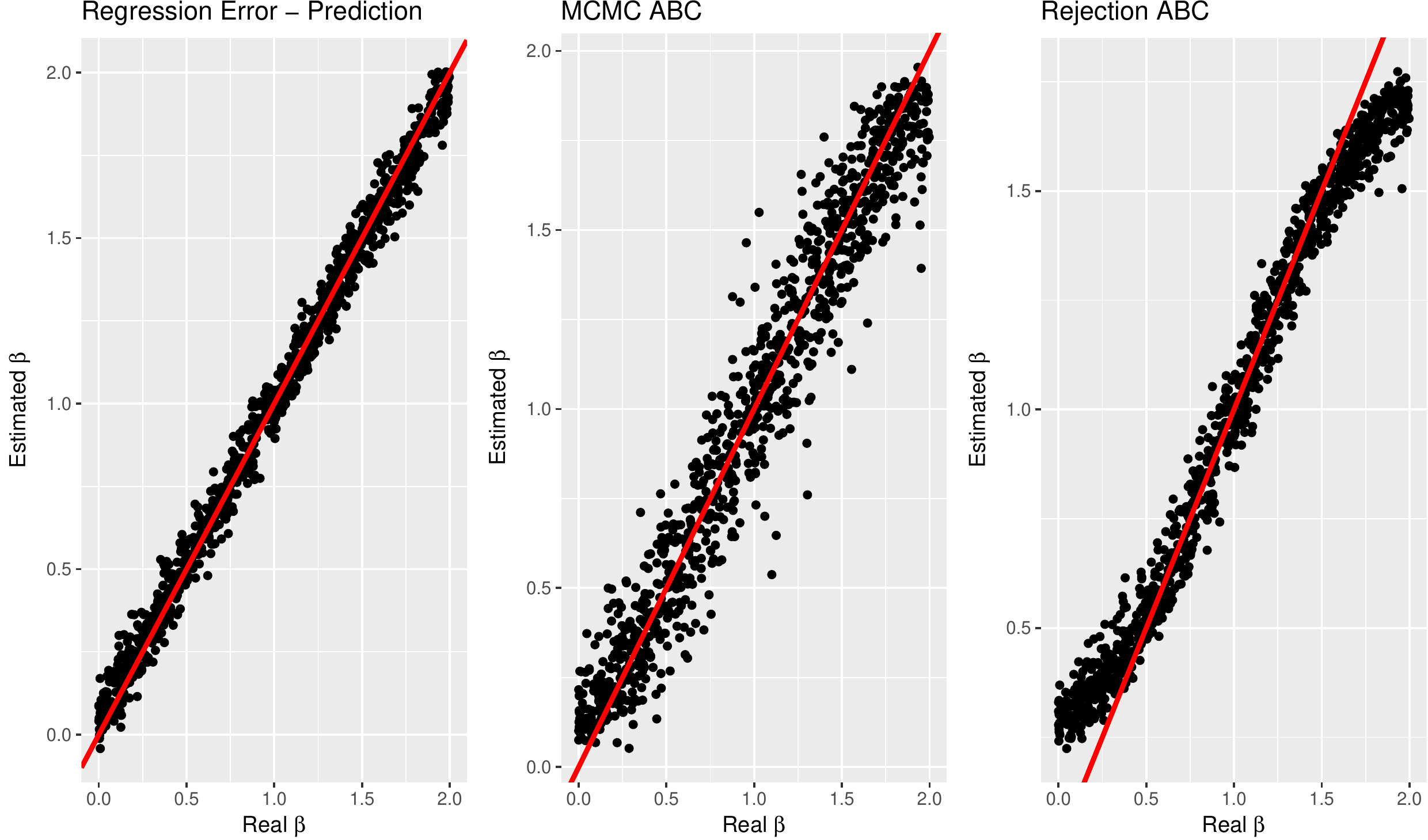}
\caption{\label{fig:brokenlinerplots}A side by side comparison between the
`real' and discovered \(\beta\) as generated by either the regression
method or two standard ABC algorithms: rejection sampling or MCMC. The
x-axis represents the real \(\beta\), the y-axis its estimate. Each
point represents the estimation against a different set of summary
statistics. The red line is the 45 degree line: the closer the points
are to it the better the estimation. Notice that rejection sampling
(ABC) is inaccurate when \(\beta\) is very small (near 0) or very large
(near 2) while MCMC has a larger average error than regression-based
methods}
\end{figure}

We run the model another 1000 times to test the regression ability to
estimate their \(\beta\). As shown in Figure \ref{fig:brokenlinerplots}
and Table \ref{tab:brokentable} the regression outperforms both ABC
alternatives. This is due to the elastic net ability to prune
unimportant summary statistics.

\begin{table}

\caption{\label{tab:brokenlinerestimation}Coefficient estimates after 1000 training observations; notice that all the uninformative summary statistics have been automatically dropped.}
\centering
\begin{tabular}[t]{l|r}
\hline
Term & Estimate\\
\hline
(Intercept) & 0.0360036\\
\hline
$b_5$ & 0.0189285\\
\hline
$b_6$ & 0.0202486\\
\hline
$b_7$ & 0.0222020\\
\hline
$b_8$ & 0.0288608\\
\hline
$b_9$ & 0.0273448\\
\hline
\end{tabular}
\end{table}

\begin{table}

\caption{\label{tab:brokentable}Regression errors achieve lower estimation errors in a fraction of the time. This is because they automatically ignore all summary statistics that do not inform $\beta$}
\centering
\begin{tabular}[t]{l|ccr}
\hline
Estimation & Average RMSE & Runtime & Runtime(scaled)\\
\hline
Regression Method & 0.0579736 & 0.0008367697 mins & 1.000\\
\hline
Rejection ABC & 0.1278042 & 0.9592413 mins & 1146.362\\
\hline
MCMC ABC & 0.1343737 & 6.273226 mins & 7496.956\\
\hline
\end{tabular}
\end{table}

\begin{figure}[!htb]
\centering
\includegraphics{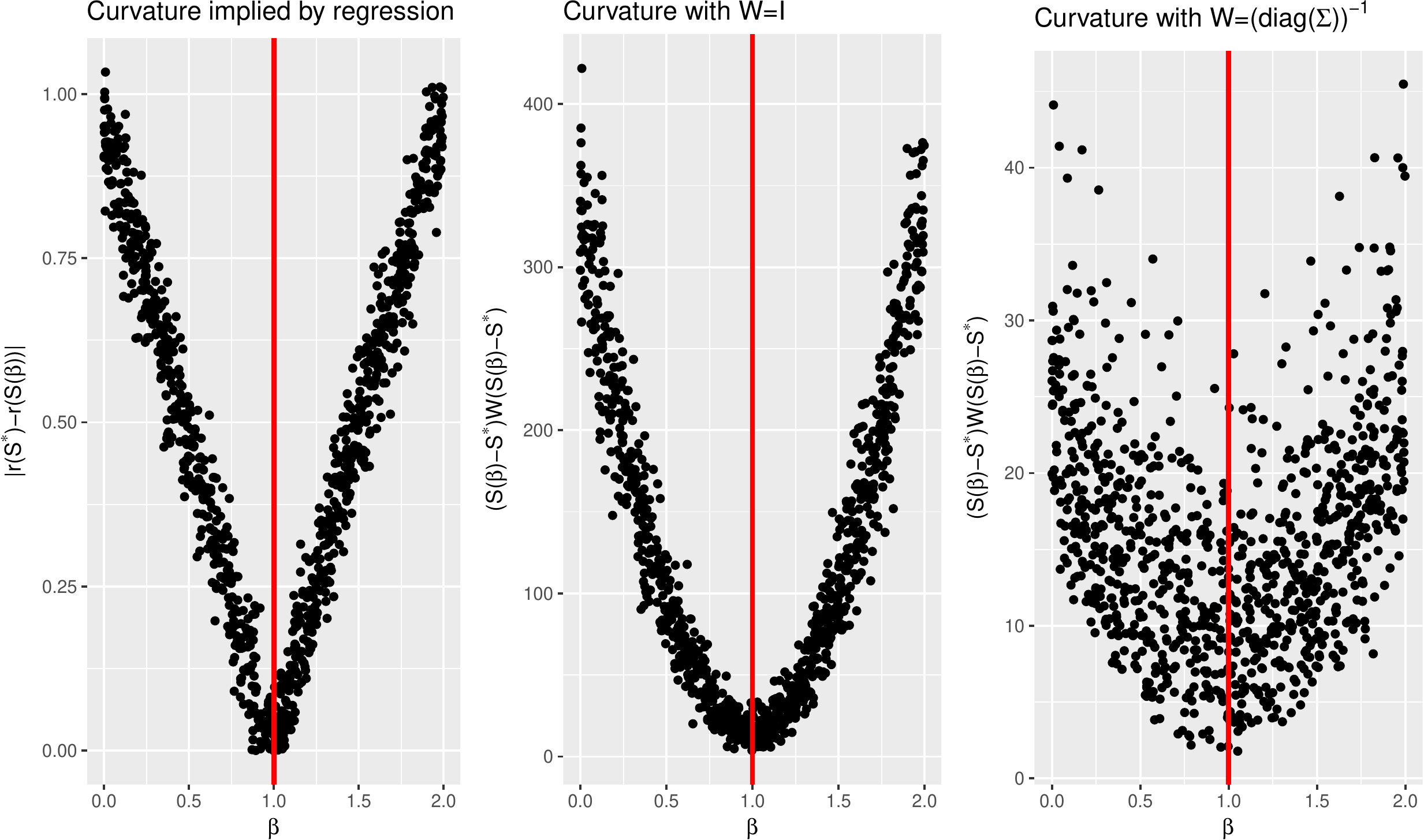}
\caption{\label{fig:brokenestimationslope}Curvature of the fitness function
as implied by the regression versus weighted Euclidean distance between
summary statistics. We take the 1000 training simulation runs, we pick
one in the middle (\(\beta \approx 1\)) as the `real' one and we compute
the summary statistics distances between all the other simulations to
it. The red line represents the real \(\beta\), each dot is the distance
to the real \(\beta\) for that simulation run and its associated
\(\beta\). For this model the default choice of weighing matrix (the
inverse of the variance diagonal) makes numerical minimization harder.}
\end{figure}

In simulated minimum distance methods, choosing the wrong \(W\) can
compound the problem of poor summary statistics selection. As shown in
Figure \ref{fig:brokenestimationslope} the default choice
\(W= \left(\text{diag}(\Sigma)\right)^{-1}\) detrimentally affects the
curvature of the distance function, by adding noise. This is because
summary statistics \(S_0,\dots,S_4\), in spite of containing no useful
information, are the values with the lowest variance and are therefore
given larger weights.

We showed here how, even in a simple one dimensional problem, the
ability to prune uninformative summary statistics will result in better
estimation (as shown by RMSE in table \ref{tab:brokentable}) .

\hypertarget{model-selection-through-regression-based-methods-is-informative-and-accurate}{%
\subsubsection{Model selection through regression-based methods is
informative and
accurate}\label{model-selection-through-regression-based-methods-is-informative-and-accurate}}

In this section we train a classifier to choose which of the two models
we described above (`straight-' and `broken-line') generated the
observed data. We show that the classifier does a better job than
choosing the model that minimizes prediction errors.

We observe 10 summary statistics \(S^*=(S^*_0,\dots,S^*_9)\) and we
would like to know if they were generated by the ``straight-line''
model: \begin{equation}
S_i = \beta i + \epsilon \label{eq:straightline} 
\end{equation} or the ``broken-line'' model: \begin{equation}
S_i=\left\{\begin{matrix}
\epsilon & i < 5\\ 
\beta i + \epsilon & i\geq5
\end{matrix}\right. \label{eq:brokenline} 
\end{equation}\\
where \(\epsilon \sim N(0,1)\) and \(i \in [0,9]\). Assume we have
already estimated both models and \(\beta=1\).

We run each model 1000 times and then train a logit classifier: \begin{equation} 
\text{Pr}(\text{broken line model}) = \frac{1}{1 + \exp(-[a +\sum b_i S_i])} 
\label{eq:brokenlineclassifier}  
\end{equation} Notice that \(S_0\) and \(S_5,\dots,S_9\) are generated by the same
rule in both models: both draw a straight line through those points.
Their values will then not be useful for model selection. We should
focus instead exclusively on \(S_1,\dots,S_4\) since that is the only
area where one model draws a straight line and the other does not.\\
The regularized regression discovers this, by dropping \(b_0\) and
\(b_5,\dots,b_9\) as Table \ref{tab:modelselectionlinetabb} shows.

\begin{table}

\caption{\label{tab:modelselectionlinetabb}Coefficient estimates after 2000 training observations; notice how only the low summary statistics have a coefficient associated to them. This is because they are the only part of the line that is 'broken': $S_5,\dots,S_9$ are generated the same way for both models and are therefore of no use for differentiating the two models.}
\centering
\begin{tabular}[t]{l|r}
\hline
Term & Estimate\\
\hline
(Intercept) & -11.1578231\\
\hline
$b_1$ & 0.5733553\\
\hline
$b_2$ & 1.1016989\\
\hline
$b_3$ & 2.0715734\\
\hline
$b_4$ & 1.8777375\\
\hline
\end{tabular}
\end{table}

\begin{figure}[!htb]
\centering
\includegraphics{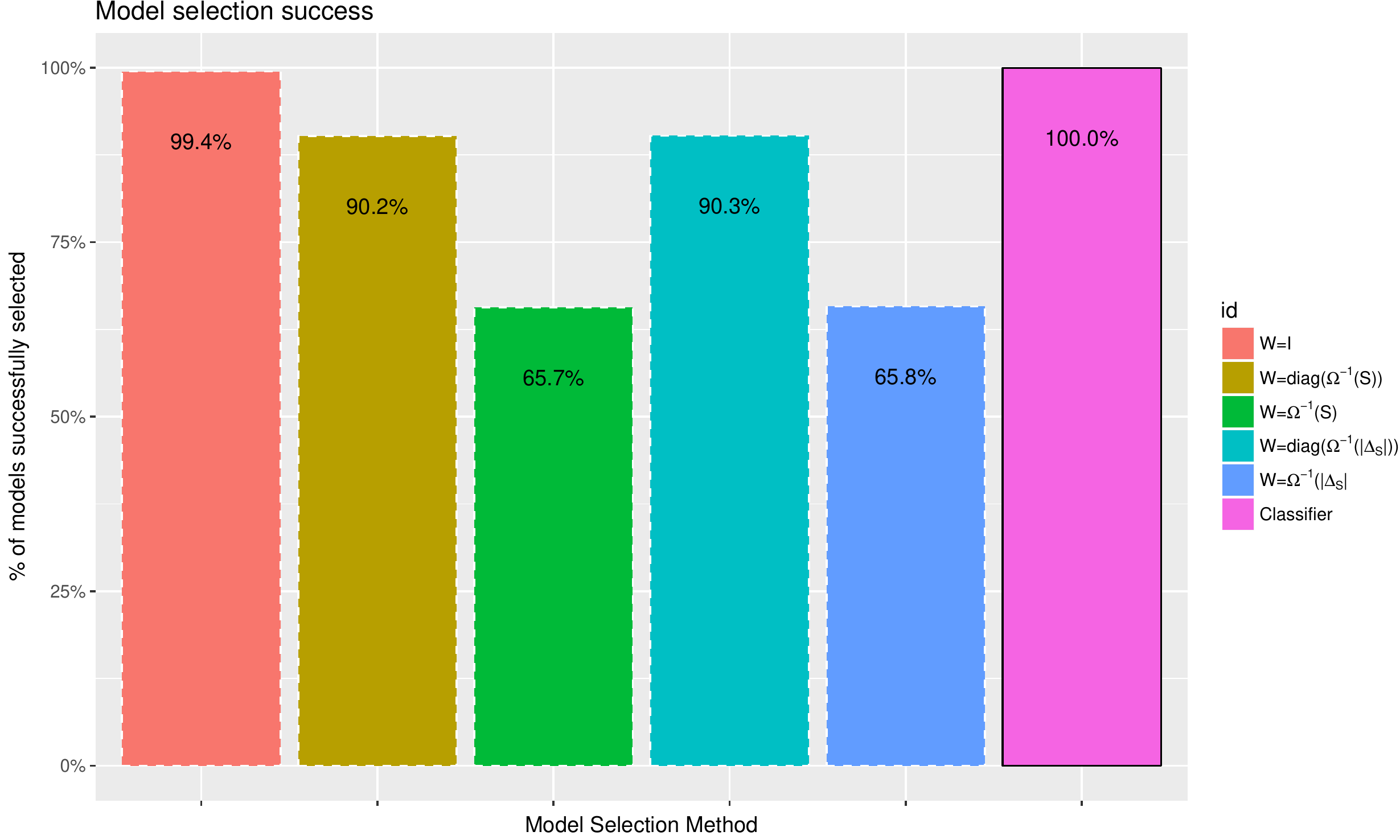}
\caption{\label{fig:modelselectionline}A comparison between training a
classifier to do model selection versus choosing the model whose
distance to summary statistics is smaller. All weighing schemes except
\(W=I\) underperform compared to the classifier. Model selection
performed on 2000 testing runs (not used for training the classifier)}
\end{figure}

Figure \ref{fig:modelselectionline} tabulates the success rate from 2000
out-of-sample simulations by the classifier and compares it with picking
models by choosing the one that minimizes the distance between summary
statistics. The problem is simple enough that choosing \(W=I\) achieves
almost perfect success rate; using the more common variance-based
weights does not.

\hypertarget{rbcexample}{%
\subsection{Fit RBC}\label{rbcexample}}

\hypertarget{crosscorrelation}{%
\subsubsection{Fit against cross-correlation
matrix}\label{crosscorrelation}}

In this section we parametrize a simple macro-economic model by looking
at the time series it simulates. We show that we can use a large number
of summary statistics efficiently and that we can diagnose estimation
strengths and weaknesses the same way we diagnose regression outputs.

We observe 150 steps\footnote{This is the setup in the original indirect
  inference paper by Smith (\protect\hyperlink{ref-Smith1993}{1993})
  applied to a similar RBC model} of 5 quarterly time series:
\(Y,r,I,C,L\). Assuming we know they come from a standard real business
cycle (RBC) model (the default RBC implementation of the gEcon package
in R, see Klima, Podemski, and Retkiewicz-Wijtiwiak
\protect\hyperlink{ref-Klima2018}{2018}, see also Appendix A), we
estimate the 6 parameters (\(\beta,\delta,\eta,\mu,\phi,\sigma\)) that
generated them.

Summarize the time series observed with (a) the \(t=-5,\dots,+5\)
cross-correlation vectors of \(Y\) against \(r,I,C,L\), (b) the
lower-triangular covariance matrix of \(Y,r,I,C,L\), (c) the squares of
all cross-correlations and covariances, (d) the pair-wise products
between all cross-correlations and covariances. This totals to 2486
summary statistics: too many for most ABC approaches.\\
This problem is however still solvable by regression. We train 6
separate regressions, one for each parameter, as follows: \begin{equation} 
\left\{\begin{matrix}
\beta &= \sum a_{\beta,i} S_i + \sum b_{\beta,i} S_{i}^2 + \sum c_{\beta,i,j} S_i S_j  \\
\gamma &= \sum a_{\gamma,i} S_i + \sum b_{\gamma,i} S_{i}^2 + \sum c_{\gamma,i,j} S_i S_j \\
&\vdots \\
\sigma &= \sum a_{\sigma,i} S_i + \sum b_{\sigma,i} S_{i}^2 + \sum c_{\sigma,i,j} S_i S_j \\
\end{matrix}\right.
\label{eq:regressioncross}  
\end{equation} Where \(a,b,c\) are elastic net coefficients.

\begin{figure}[!htb]
\centering
\includegraphics{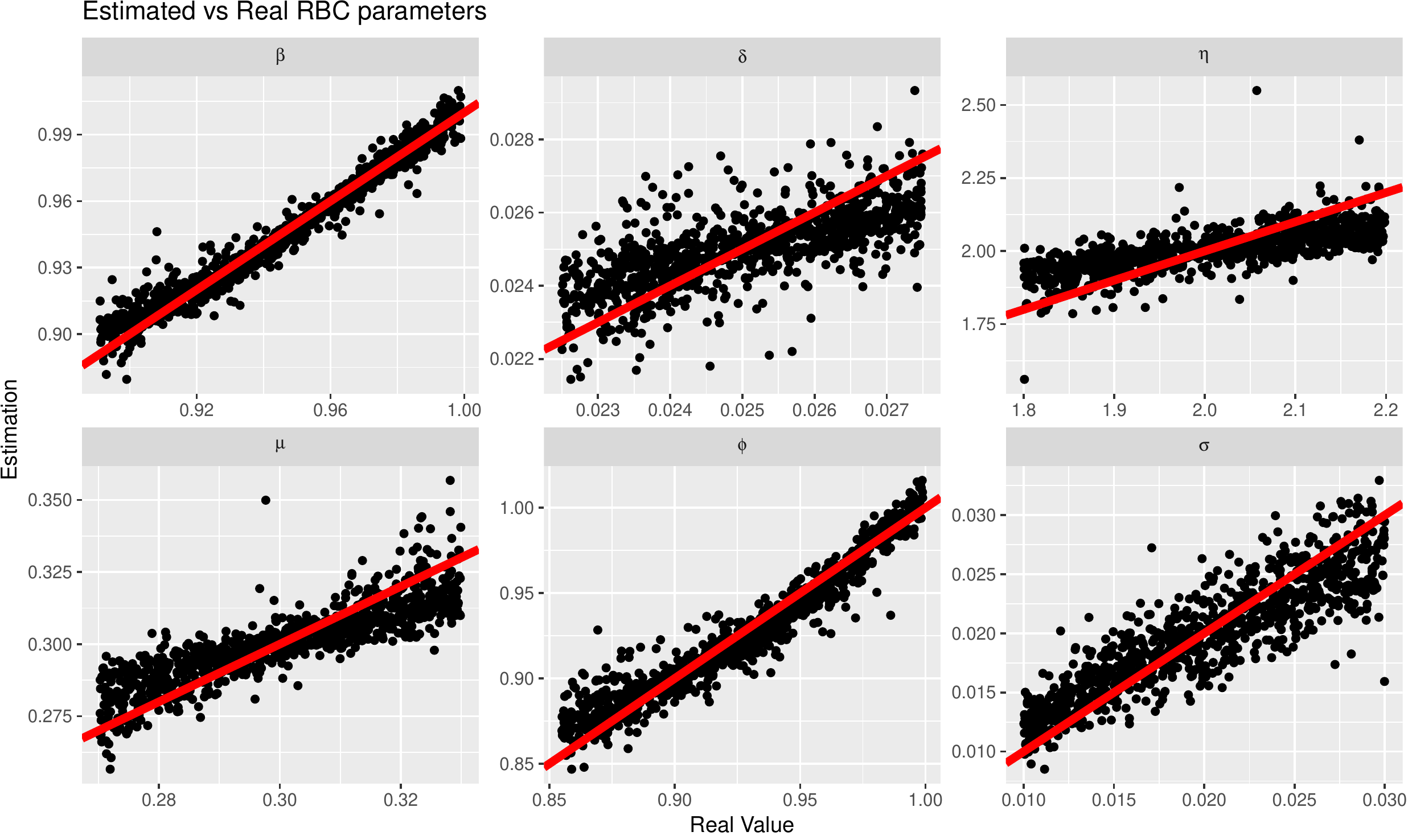}
\caption{\label{fig:rbccovplott}Estimated versus real parameters for 2000
RBC runs in the testing data-set (not used for training the
regressions). Each dot is a pair real-estimate parameter from a single
RBC run. Red line is the 45 degree line: the closer the dots are to it
the better estimated that parameter is. The parameters \(\delta\) and
\(\eta\) are not as well estimated as the other 4}
\end{figure}

We train the regressions over 2000 RBC runs then produce 2000 more to
test its estimation. Figure \ref{fig:rbccovplott} and Table
\ref{tab:rbccovtable} show that all the parameters are estimated
correctly but \(\eta\) and \(\delta\) less precisely than the others. We
judge the quality of the estimation through the predictivity coefficient
(see Salle and Yıldızoğlu \protect\hyperlink{ref-Salle2014}{2014}; also
known as modelling efficiency as in Stow et al.
\protect\hyperlink{ref-Stow2009}{2009}) which measures the improvement
of using an estimate \(\hat \beta\) to discover the real \(\beta^*\) as
opposed to just using the hindsight average \(\bar \beta\): \begin{equation}
\text{Predictivity} = \frac{ \sum \left( \beta^*_i - \bar \beta \right)^2 - \sum \left( \beta^*_i - \hat \beta_i \right)^2 }{\sum \left( \beta^*_i - \bar \beta \right)^2}
\label{eq:predictivity}  
\end{equation} Intuitively, this is just an extension of the definition to \(R^2\)
to out-of-sample predictions. A predictivity of 1 implies perfect
estimation, 0 implies that averaging would have performed as well as the
estimate.

\begin{table}
\caption{\label{tab:rbccovtable}Predictivity for each
variable when estimating parameters of the RBC model looking at
cross-correlations and covariance; \(\delta\) and \(\eta\) are less
precisely estimated than the other parameters. All results obtained on a
testing data-set (not used for training) of 2000 RBC runs. Bias is
defined as average distance between real and estimated parameter:
\(\theta^*-\theta\), RMSE is the root mean square error:
\(\sqrt{\sum \frac{\left(\theta^*-\theta\right)^2}{2000}}\). Variable
bounds are the interval 20\% below and above the default RBC parameters
in the original gEcon implementation.} 
\centering
\begin{tabular}[t]{l|l|r|r|r}
\hline
Variable & Variable Bounds & Average Bias & Average RMSE & Predictivity\\
\hline
$\beta$ & [0.891,0.999] & 0.0003105 & 0.0000351 & 0.9635677\\
\hline
$\delta$ & [0.225,0.275] & -0.0000426 & 0.0000010 & 0.4997108\\
\hline
$\eta$ & [1.8,2.2] & -0.0009051 & 0.0062672 & 0.5168588\\
\hline
$\mu$ & [0.27,0.33] & 0.0001315 & 0.0000774 & 0.7419450\\
\hline
$\sigma$ & [0.01,0.03] & -0.0000391 & 0.0000071 & 0.7941670\\
\hline
$\phi$ & [0.855,0.999] & -0.0001568 & 0.0001203 & 0.9265618\\
\hline
\end{tabular}
\end{table}

We showed here that we can estimate the parameters of a simple RBC model
by looking at their cross-correlations. However this method is better
served by looking at multiple auxiliary models as we show in the next
section.

\hypertarget{fit-against-auxiliary-fits}{%
\subsubsection{Fit against auxiliary
fits}\label{fit-against-auxiliary-fits}}

In this section we parametrize the same simple macro-economic model by
looking at the same time series it simulates but using a more diverse
array of summary statistics. We show how that improves the estimation of
the model.

Again, we observe 150 steps of 5 quarterly time series: \(Y,r,I,C,L\).
Assuming we know they come from a standard RBC model, we want to
estimate the 6 parameters (\(\beta,\delta,\eta,\mu,\phi,\sigma\)) that
generated them.

In this section we use a larger variety of summary statistics. Even
though they derive from auxiliary models that convey much of the same
information, we are able to use them efficiently for estimation. We
summarize output by (i) coefficients of regressing \(Y\) on
\(Y_{t-1},I_{t},I_{t-1}\), (ii) coefficients of regressing \(Y\) on
\(Y_{t-1},C_{t},C_{t-1}\), (iii) coefficients of regressing \(Y\) on
\(Y_{t-1},r_{t},r_{t-1}\), (iv) coefficients of regressing \(Y\) on
\(Y_{t-1},L_{t},L_{t-1}\), (v) coefficients of regressing \(Y\) on
\(C,r\) (vi) coefficients of fitting AR(5) on \(Y\), (vii) the (lower
triangular) covariance matrix of \(Y,I,C,r,L\).\\
In total there are 48 summary statistics to which we add all their
squares and cross-products.

\begin{figure}[!htb]
\centering
\includegraphics{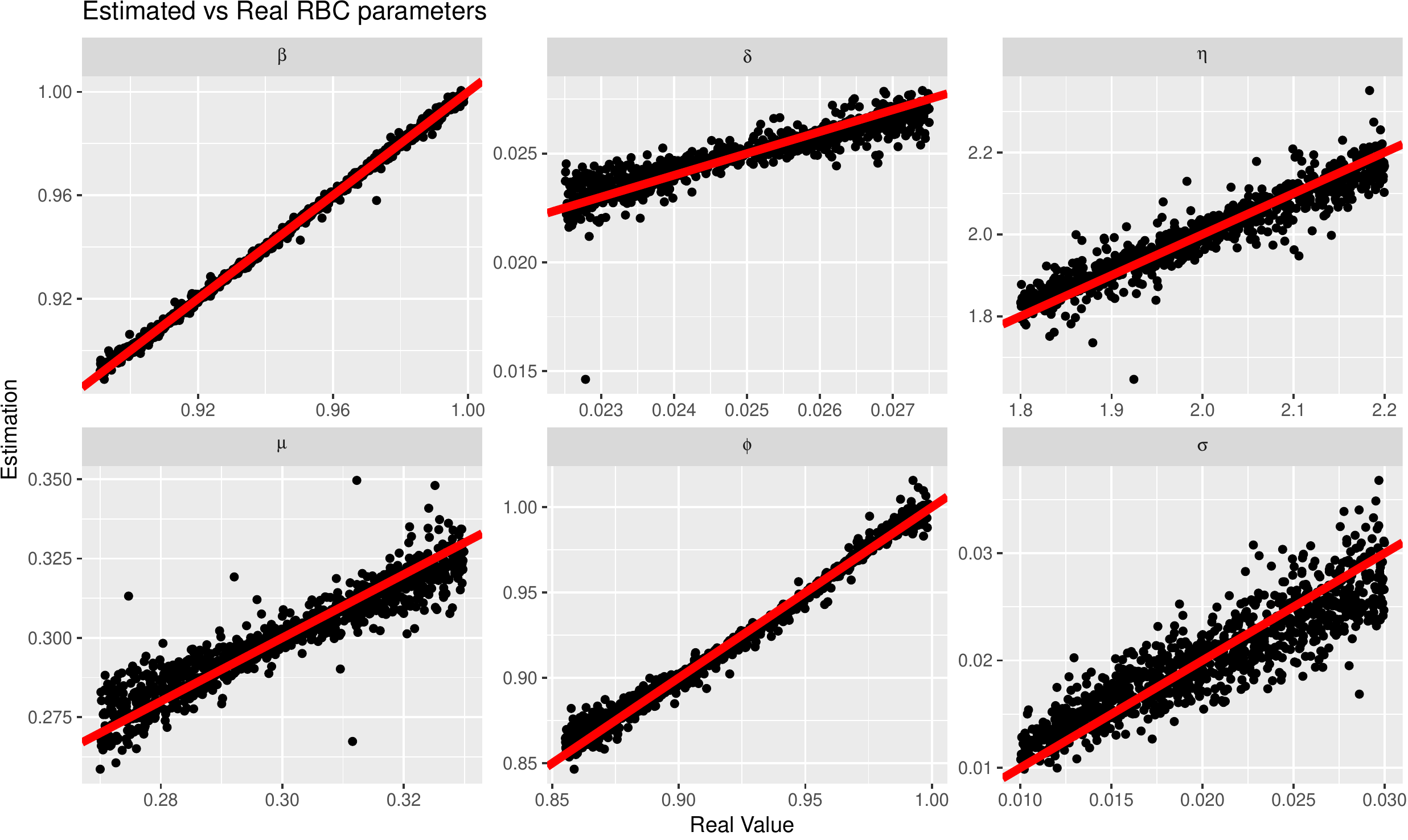}
\caption{\label{fig:rbcfullplot}Estimated versus real parameters for 2000
RBC runs in the testing data-set (not used for training the
regressions). Each dot is a pair of real-estimate parameters from a
single RBC run. The red line is the 45 degree line: the closer the dots
are to it the better estimated that parameter is. All parameters are
well estimated}
\end{figure}

Again we train 6 regressions against 2000 RBC observations and test
their estimation against 2000 more.\\
Figure \ref{fig:rbcfullplot} and Table \ref{tab:rbcfulltable} shows that
all the parameters are estimated correctly and with higher predictivity
coefficient. In particular \(\delta\) and \(\eta\) are better predicted
than by looking at cross-correlations.

\begin{table}
\caption{\label{tab:rbcfulltable}Predictivity for each
variable when estimating parameters of the RBC model looking at
cross-correlations and covariance; All parameters are well estimated.
All results obtained on a testing data-set (not used for training) of
2000 RBC runs. Bias is defined as average distance between real and
estimated parameter: \(\theta^*-\theta\), RMSE is the root mean square
error: \(\sqrt{\sum \frac{\left(\theta^*-\theta\right)^2}{2000}}\).
Variable bounds are the interval 20\% below and above the default RBC
parameters in the original gEcon implementation.} 
\centering
\begin{tabular}[t]{l|l|r|r|r}
\hline
Variable & Variable Bounds & Average Bias & Average RMSE & Predictivity\\
\hline
$\beta$ & [0.891,0.999] & 0.0000154 & 0.0000023 & 0.9975601\\
\hline
$\delta$ & [0.225,0.275] & -0.0000308 & 0.0000003 & 0.8285072\\
\hline
$\eta$ & [1.8,2.2] & 0.0001992 & 0.0012552 & 0.9040992\\
\hline
$\mu$ & [0.27,0.33] & 0.0000126 & 0.0000357 & 0.8793255\\
\hline
$\sigma$ & [0.01,0.03] & -0.0000638 & 0.0000061 & 0.8100463\\
\hline
$\phi$ & [0.855,0.999] & 0.0002740 & 0.0000244 & 0.9862208\\
\hline
\end{tabular}
\end{table}

\hypertarget{abmexample}{%
\subsection{Heuristic selection in the face of input
uncertainty}\label{abmexample}}

In this section we select between 10 variations of an agent-based model
where some inputs and parameters are unknowable. By training a
classifier we are able to discover which summary statistics are robust
to the unknown parameters and leverage this to outperform standard model
selection by prediction error.

Assume all fishermen in the world can be described by one of 10
decision-making algorithms (heuristics). We observe 100 fishermen for
five years and we want to identify which heuristic they are using. The
heuristic selection is however complicated by not knowing basic
biological facts about the environment the fishermen exploit.\\
This is a common scenario in fisheries: we can monitor boats precisely
through electronic logbooks and satellites but more than 80\% of global
catch comes from unmeasured fish stocks(Costello et al.
\protect\hyperlink{ref-Costello2012}{2012}).\\
Because heuristics are adaptive, the same heuristic will generate very
different summary statistics when facing different biological
constraints. We show here that we can still identify heuristics even
without knowing how many fish there are in the sea.

We can treat each heuristic as a separate model and train a classifier
that looks at summary statistics to predict which heuristic generated
them. The critical step is to train the classifier on example runs where
the uncertain inputs (the unknown biological constraints) are
randomized. This way the classifier automatically learns which summary
statistics are robust to biological changes and look only at those when
selecting the right heuristic.

We use POSEIDON(Bailey et al. \protect\hyperlink{ref-Bailey2018}{2018}),
a fishery agent-based model to simulate the fishery. We observe the
fishery for five years and we collect 17 summary statistics: (i)
aggregate five year averages on landings, effort (hours spent fishing),
distance travelled from port, number of trips per year and average hours
at sea per trip, (ii) a discretized(3 by 3 matrix) heatmap of
geographical effort over the simulated ocean, (iii) the coefficients and
\(R^2\) of a discrete choice model fitting area fished as a function of
distance from port and habit (an integer describing how many times that
particular fisher has visited that area before in the past 365 days ).\\
These summary statistics are a typical information set we might expect to obtain in a developed fishery.

Within each simulation, all fishers use the same heuristic. There are 10
possible candidates,all described more in depth in a separate
paper\footnote{currently in R\&R, working draft:
  \url{http://carrknight.github.io/poseidon/algorithms.html}}:

\begin{itemize}
\tightlist
\item
  Random: each fisher each trip picks a random cell to trawl
\item
  Gravitational Search: agents copy one another as if planets attracted
  by gravity where mass is equal to profits made in the previous trip
  (see Rashedi, Nezamabadi-pour, and Saryazdi
  \protect\hyperlink{ref-rashedi_gsa:_2009}{2009})
\item
  \(\epsilon\)-greedy bandit: agents have a fixed 20\% chance of
  exploring a new random area otherwise exploiting(fishing) the
  currently most profitable discovered location. Two parametrizations of
  this heuristic are available, depending on whether fishers discretize
  the map in a 3 by 3 matrix or a 9 by 9 one.
\item
  ``Perfect'' agent: agents knows perfectly the profitability of each
  area \(\Pi(\cdot)\) and chooses to fish in area \(i\) by SOFTMAX: \begin{equation}
  \text{Probability}(i) = \frac{e^{\Pi(i)}}{\sum_j e^{\Pi(j)}}  \label{eq:perfectagents}  
  \end{equation}
\item
  Social Annealing: agents always return to the same location unless
  they are making less than fishery's average profits(across the whole
  fishery), in which case they try a new area at random
\item
  Explore-Exploit-Imitate: like \(\epsilon\)-greedy agents, they have a
  fixed probability of exploring. When not exploring each fisher checks
  the profits of two random ``friends'' and copies their location if
  either are doing better. Three parametrizations of this heuristics are
  possible: exploration rate at 80\% or 20\% with exploration range of 5
  cells, or 20\% exploration rate with exploration range of 20 cells.
\item
  Nearest Neighbour: agents keep track of the profitability of all the
  areas they have visited through a nearest neighbour regression and
  always fish the peak area predicted by the regression.
\end{itemize}

All runs are simulations of the single-species ``baseline'' fishery
described in the ESM of Bailey et al.
(\protect\hyperlink{ref-Bailey2018}{2018}) (see also Appendix B).\\
We run the model 10,000 times in total, 1000 times for each heuristic.
We also randomize in each simulation three biological parameters: (i)
carrying capacity of the each cell map, (ii) diffusion speed of fish,
(iii) the Malthusian growth parameter of the stock.

We train a multi-logit classifier on the simulation outputs as shown in
section \ref{patternrecognition}: \begin{equation} 
\text{Pr}(\text{Heuristic }i) = \frac{e^{b_i' S}}{1+\sum_j e^{b_j' S}} \label{eq:multiclassifier}
\end{equation} where \(b_1,\dots,b_{10}\) are vectors of elastic net coefficients.

\begin{figure}[!htb]
\centering
\includegraphics{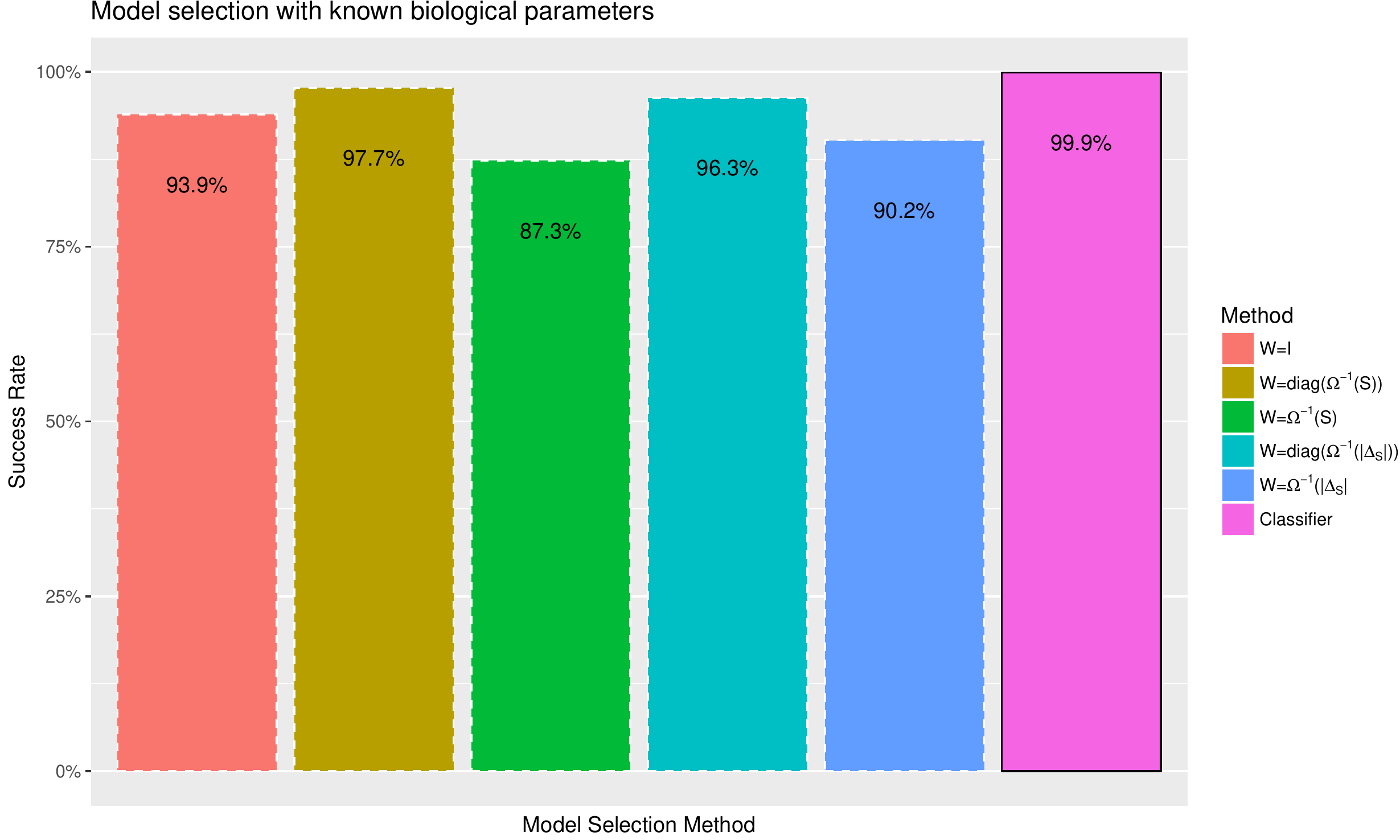}
\caption{\label{fig:easyclassification}Comparison of success rate in
correctly estimating which heuristics generated a given summary
statistics \(S\). The classifier success rate has solid borders. In this
set of simulations model selection by minimum prediction error is
advantaged by knowing the correct biological parameters while the
classifier is a single global model trained over all combinations of
biological parameters.}
\end{figure}

\begin{figure}[!htb]
\centering
\includegraphics{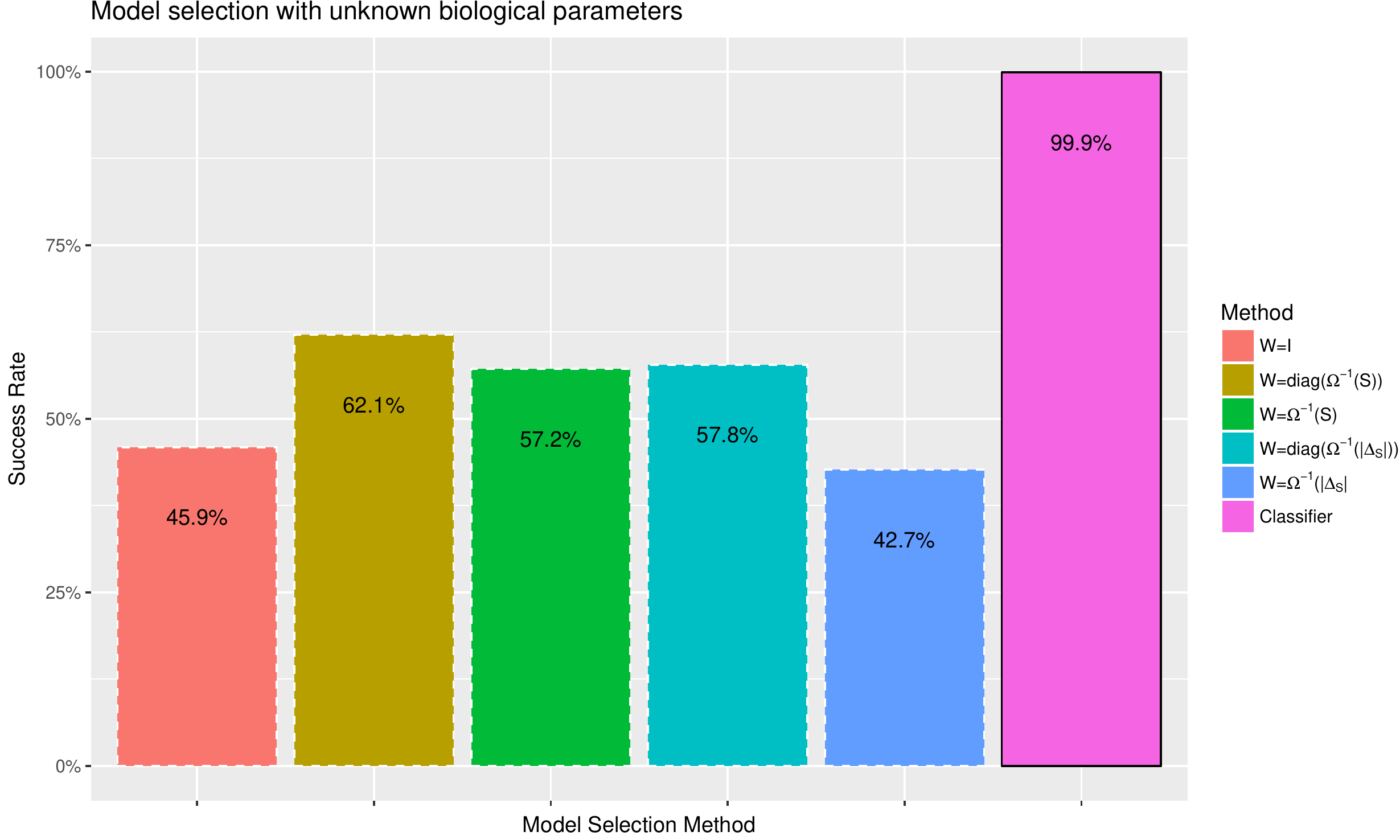}
\caption{\label{fig:abmerrorcomparison}Comparison of success rate in
correctly estimating which heuristics generated a given summary
statistics \(S\). The classifier success rate has solid borders. In this
set of simulations model selection by minimum prediction error has no
additional knowledge about the biological parameters.}
\end{figure}

We build the testing data-set by running the model 10,000 more times. We
compare the quality of model selection against selection by weighted
prediction error.\\
First, in Figure \ref{fig:easyclassification} the prediction error is
computed by re-running simulations with the correct random biological
parameters. In spite of having a knowledge advantage over the
classifier, selection by prediction error performs worse than the
classifier. In Figure \ref{fig:abmerrorcomparison} we do not assume to
know the random biological parameters. Prediction error (which now tries
to minimize the distance between simulations which differ in both
heuristics and biological parameters) performs much worse.

Model selection by prediction error is hard if we are not certain about
all model inputs. However it is trivial to train a classifier by feeding
it observations from different parameter configurations such that only
the summary statistics ``robust'' to such parameter noise are used for
model selection.

\hypertarget{concludes}{%
\section{Discussion}\label{concludes}}

\hypertarget{advantages-of-a-regression-based-approach}{%
\subsection{Advantages of a regression-based
approach}\label{advantages-of-a-regression-based-approach}}

Regression are commonly-used and well-understood. We leverage that
knowledge for parameter estimation.\\
By looking at out-of-sample prediction we can immediately diagnose
parameter identification problems: the lower the accuracy, the weaker
the identification (compare this to diagnosis of numerical minimization
outputs as in Canova and Sala \protect\hyperlink{ref-Canova2005}{2009})
. Moreover, if the regression is linear, its coefficients show which
summary statistic \(S_i\) inform which model parameter \(\theta_i\).

A secondary advantage of building a statistical model is the ease with
which additional input uncertainty can be incorporated as shown in
section \ref{abmexample}. If we are uncertain about an input and cannot
estimate it directly (because of lack of data or under-identification),
it is straightforward to simulate for multiple values of the input,
train a statistical model and then check whether it is possible to
estimate reliably the other parameters. In a simulated minimum distance
setting we would have to commit to a specific guess of the uncertain
input before running a minimization or minimize multiple times for each
guess (without any assurance that the weighing matrix ought to be kept
constant between guesses).

\hypertarget{weaknesses}{%
\subsection{Weaknesses}\label{weaknesses}}

This method has two main limits. First, users cannot weight summary
statistics by importance. As a consequence, the method cannot be
directed to replicate a specific summary statistic. Imagine a financial
model of the housing market which we may judge primarily by its ability
of matching real house prices. However, when estimating its parameters
the regression may ignore house prices entirely and focus on another
seemingly minor summary statistic (say, total number of bathrooms).
While that may be an interesting and useful way of parametrizing the
model, it does not guarantee a good match with observed house prices.\\
Nevertheless, in this case one option would be to transition to a
weighted regression and weigh more the observations whose generated
summary statistic \(S_i\) were close to the value of \(S_i^*\) we are
primarily interested in replicating. As such, the model would be forced
to include component deemed important, even if not the most efficient in
terms of explanatory power. Comparing this fit with the unweighted one
may also be instructive.

The second limit is that the each parameter is estimated separately.
Because each parameter is given a separate regression, this method may
underestimate underlying connections (dependencies) between two or more
parameters when such connections are not reflected in changes to summary
statistics. This issue is mentioned in the ABC literature by Beaumont
(\protect\hyperlink{ref-Beaumont2010}{2010}) although it is also claimed
that in practice it does not seem to be an issue. It may however be more
appropriate to use a simultaneous system estimation routine(Henningsen
and Hamann \protect\hyperlink{ref-Henningsen2007}{2007}) or a multi-task
learning algorithm(Evgeniou and Pontil
\protect\hyperlink{ref-Evgeniou2004}{2004}).

\hypertarget{possible-extensions}{%
\subsection{Possible Extensions}\label{possible-extensions}}

Our method contains a fundamental inefficiency: we build a global
statistical model to use only once when plugging in the real summary
statistics \(S^*\). Given that the real summary statistics \(S^*\) are
known in advance, a direct search ought to be more efficient.\\
Our method makes up for this by solving multiple problems at once as it
implicitly selects summary statistics and weights them while estimating
the model parameters.

In machine learning, transduction substitutes regression when we are
interested in a prediction for a known \(S^*\) only, without the need to
build a global statistical model(see Chapter 10 of Cherkassky and Mulier
\protect\hyperlink{ref-Cherkassky2007}{2007}; Beaumont, Zhang, and
Balding \protect\hyperlink{ref-Beaumont2002}{2002} weights simulations
depending on the distance to known \(S^*\) in essentially the same
approach). Switching to transduction would however remove the
familiarity advantage of using regressions as well as the ease to test
its predictive power with randomly generated summary statistics.

Another avenue for improvement is to use non-linear, non-parametric
regressions. Blum and Francois (\protect\hyperlink{ref-Blum2010}{2010})
used neural networks, for example. Non-parametrics are more flexible but
they may be more computationally demanding and their final output harder
to interpret.

It may also be possible to use regressions as an intermediate step in a
more traditional simulated minimum distance problem. We would still
build a regression for each parameter but then using them as the
distance function to minimize by standard numerical methods.

\hypertarget{conclusion}{%
\subsection{Conclusion}\label{conclusion}}

We presented here a general method to fit agent-based and simulation
models. We perform indirect inference using prediction rather than
minimization, and, by using regularized regressions, we avoid three
major problems of estimation: selecting the most valuable of the summary
statistics, defining the distance function, and achieving reliable
numerical minimization. By substituting regression with classification
we can further extend this approach to model selection. This approach is
relatively easy to understand and apply, and at its core uses
statistical tools already familiar to social scientists. The method
scales well with the complexity of the underlying problem, and as such
may find broad application.

\hypertarget{appendix-a-rbc-model}{%
\section*{Appendix A: RBC Model}\label{appendix-a-rbc-model}}
\addcontentsline{toc}{section}{Appendix A: RBC Model}

\hypertarget{consumer}{%
\subsection*{Consumer}\label{consumer}}
\addcontentsline{toc}{subsection}{Consumer}

\begin{align}
&\max_{K^{\mathrm{s}}_{t}, C_{t}, L^{\mathrm{s}}_{t}, I_{t}
} U_{t} = {\beta} {\mathrm{E}_{t}\left[U_{t+1}\right]} + \left(1 - \eta\right)^{-1} {\left({{C_{t}}^{\mu}} {\left(1 - L^{\mathrm{s}}_{t}\right)^{1 - \mu}}\right)^{1 - \eta}}\\
& C_{t} + I_{t} = \pi_{t} + {K^{\mathrm{s}}_{t-1}} {r_{t}} + {L^{\mathrm{s}}_{t}} {W_{t}} \\
& K^{\mathrm{s}}_{t} = I_{t} + {K^{\mathrm{s}}_{t-1}} \left(1 - \delta\right)
\end{align}

\hypertarget{firm}{%
\subsection*{Firm}\label{firm}}
\addcontentsline{toc}{subsection}{Firm}

\begin{align}
&\max_{K^{\mathrm{d}}_{t}, L^{\mathrm{d}}_{t}, Y_{t}
} \pi_{t} = Y_{t} - {L^{\mathrm{d}}_{t}} {W_{t}} - {r_{t}} {K^{\mathrm{d}}_{t}}\\
& Y_{t} = {Z_{t}} {{K^{\mathrm{d}}_{t}}^{\alpha}} {{L^{\mathrm{d}}_{t}}^{1 - \alpha}} 
\end{align}

\hypertarget{equilibrium}{%
\subsection*{Equilibrium}\label{equilibrium}}
\addcontentsline{toc}{subsection}{Equilibrium}

\begin{equation}
K^{\mathrm{d}}_{t} = K^{\mathrm{s}}_{t-1} 
\end{equation} \begin{equation}
L^{\mathrm{d}}_{t} = L^{\mathrm{s}}_{t}
\end{equation} \begin{equation}
Z_{t} = e^{\epsilon^{\mathrm{Z}}_{t} + {\phi} {\log{Z_{t-1}}}}
\end{equation} \begin{equation}
-0.36Y_\mathrm{ss} + {r_\mathrm{ss}} {K^{\mathrm{s}}_\mathrm{ss}} = 0
\end{equation}

\hypertarget{appendix-b-fishery-parameters}{%
\section*{Appendix B: Fishery
Parameters}\label{appendix-b-fishery-parameters}}
\addcontentsline{toc}{section}{Appendix B: Fishery Parameters}

\begin{longtable}[]{@{}lcr@{}}
\toprule
Parameter & Value & Meaning\tabularnewline
\midrule
\endhead
\textbf{Biology} & Logistic &\tabularnewline
\(K\) & \(\sim U[1000,10000]\) & max units of fish per
cell\tabularnewline
\(m\) & \(\sim U[0,0.003]\) & fish speed\tabularnewline
\(r\) & \(\sim U[.3,.8]\) & Malthusian growth parameter\tabularnewline
\textbf{Fisher} & Explore-Exploit-Imitate &\tabularnewline
rest hours & 12 & rest at ports in hours\tabularnewline
\(\epsilon\) & 0.2 & exploration rate\tabularnewline
\(\delta\) & \(\sim U[1,10]\) & exploration area size\tabularnewline
fishers & 100 & number of fishers\tabularnewline
friendships & 2 & number of friends each fisher has\tabularnewline
max days at sea & 5 & time after which boats must come
home\tabularnewline
\textbf{Map } & &\tabularnewline
width & 50 & map size horizontally\tabularnewline
height & 50 & map size vertically\tabularnewline
port position & 40,25 & location of port\tabularnewline
cell width & 10 & width (and height) of each cell map\tabularnewline
\textbf{Market} & &\tabularnewline
market price & 10 & \$ per unit of fish sold\tabularnewline
gas price & 0.01 & \$ per litre of gas\tabularnewline
\textbf{Gear} & &\tabularnewline
catchability & 0.01 & \% biomass caught per tow hour\tabularnewline
speed & 5.0 & distance/h of boat\tabularnewline
hold size & 100 & max units of fish storable in boat\tabularnewline
litres per unit of distance & 10 & litres consumed per distance
travelled\tabularnewline
litres per trawling hour & 5 & litres consumed per hour
trawled\tabularnewline
\bottomrule
\end{longtable}

\hypertarget{references}{%
\section*{References}\label{references}}
\addcontentsline{toc}{section}{References}

\hypertarget{refs}{}
\leavevmode\hypertarget{ref-Altonji1996}{}%
Altonji, Joseph G., and Lewis M. Segal. 1996. ``Small-Sample Bias in GMM
Estimation of Covariance Structures.'' \emph{Journal of Business \&
Economic Statistics} 14 (3). Taylor \& Francis, Ltd.American Statistical
Association:353. \url{https://doi.org/10.2307/1392447}.

\leavevmode\hypertarget{ref-Badham2017}{}%
Badham, Jennifer, Chipp Jansen, Nigel Shardlow, and Thomas French. 2017.
``Calibrating with Multiple Criteria: A Demonstration of Dominance.''
\emph{Journal of Artificial Societies and Social Simulation} 20 (2).
JASSS:11. \url{https://doi.org/10.18564/jasss.3212}.

\leavevmode\hypertarget{ref-Bailey2018}{}%
Bailey, R, Ernesto Carrella, Robert Axtell, M Burgess, R Cabral, Michael
Drexler, Chris Dorsett, J Madsen, Andreas Merkl, and Steven Saul. 2018.
``A computational approach to managing coupled human-environmental
systems: the POSEIDON model of ocean fisheries.'' \emph{Sustainability
Science}, June. Springer Japan, 1--17.
\url{https://doi.org/10.1007/s11625-018-0579-9}.

\leavevmode\hypertarget{ref-Barde2017}{}%
Barde, Sylvain. 2017. ``A Practical, Accurate, Information Criterion for
Nth Order Markov Processes.'' \emph{Computational Economics} 50 (2).
Springer US:281--324. \url{https://doi.org/10.1007/s10614-016-9617-9}.

\leavevmode\hypertarget{ref-Beaumont2010}{}%
Beaumont, Mark A. 2010. ``Approximate Bayesian Computation in Evolution
and Ecology.'' \emph{Annual Review of Ecology, Evolution, and
Systematics} 41 (1):379--406.
\url{https://doi.org/10.1146/annurev-ecolsys-102209-144621}.

\leavevmode\hypertarget{ref-Beaumont2002}{}%
Beaumont, Mark A, Wenyang Zhang, and David J Balding. 2002.
``Approximate Bayesian computation in population genetics.''
\emph{Genetics} 162 (4):2025--35.
\href{https://doi.org/Genetics\%20December\%201,\%202002\%20vol.\%20162\%20no.\%204\%202025-2035}{https://doi.org/Genetics December 1, 2002 vol. 162 no. 4 2025-2035}.

\leavevmode\hypertarget{ref-Blum2013}{}%
Blum, M G B, M A Nunes, D Prangle, and S A Sisson. 2013. ``A comparative
review of dimension reduction methods in approximate Bayesian
computation.'' \emph{Statistical Science} 28 (2):189--208.
\url{https://doi.org/10.1214/12-STS406}.

\leavevmode\hypertarget{ref-Blum2010}{}%
Blum, Michael G B, and Olivier Francois. 2010. ``Non-linear regression
models for Approximate Bayesian Computation.'' \emph{Statistics and
Computing} 20 (1):63--73.
\url{https://doi.org/10.1007/s11222-009-9116-0}.

\leavevmode\hypertarget{ref-Bruins2018}{}%
Bruins, Marianne, James A Duffy, Michael P Keane, and Anthony A Smith.
2018. ``Generalized indirect inference for discrete choice models.''
\emph{Journal of Econometrics} 205 (1):177--203.
\url{https://doi.org/10.1016/j.jeconom.2018.03.010}.

\leavevmode\hypertarget{ref-Calver2006}{}%
Calver, Benoit, and Guillaume Hutzler. 2006. ``Automatic Tuning of
Agent-Based Models Using Genetic Algorithms.'' In \emph{Lecture Notes in
Computer Science}, 3891:41--57. Springer, Berlin, Heidelberg.
\url{https://doi.org/10.1007/11734680}.

\leavevmode\hypertarget{ref-Canova2005}{}%
Canova, Fabio, and Luca Sala. 2009. ``Back to square one: Identification
issues in DSGE models.'' \emph{Journal of Monetary Economics} 56
(4):431--49. \url{https://doi.org/10.1016/j.jmoneco.2009.03.014}.

\leavevmode\hypertarget{ref-Cherkassky2007}{}%
Cherkassky, Vladimir S., and Filip. Mulier. 2007. \emph{Learning from
data : concepts, theory, and methods}. IEEE Press.
\url{https://www.wiley.com/en-gb/Learning+from+Data:+Concepts,+Theory,+and+Methods,+2nd+Edition-p-9780471681823}.

\leavevmode\hypertarget{ref-Ciampaglia2013}{}%
Ciampaglia, Giovanni Luca. 2013. ``A framework for the calibration of
social simulation models.'' \emph{Advances in Complex Systems} 16
(04n05). World Scientific Publishing Company:1350030.
\url{https://doi.org/10.1142/S0219525913500306}.

\leavevmode\hypertarget{ref-Costello2012}{}%
Costello, Christopher, Daniel Ovando, Ray Hilborn, Steven D. Gaines,
Olivier Deschenes, and Sarah E. Lester. 2012. ``Status and solutions for
the world's unassessed fisheries.'' \emph{Science} 338 (6106):517--20.
\url{https://doi.org/10.1126/science.1223389}.

\leavevmode\hypertarget{ref-Drovandi2011}{}%
Drovandi, Christopher C., Anthony N. Pettitt, and Malcolm J. Faddy.
2011. ``Approximate Bayesian computation using indirect inference.''
\emph{Journal of the Royal Statistical Society. Series C: Applied
Statistics} 60 (3). Blackwell Publishing Ltd:317--37.
\url{https://doi.org/10.1111/j.1467-9876.2010.00747.x}.

\leavevmode\hypertarget{ref-Evgeniou2004}{}%
Evgeniou, Theodoros, and Massimiliano Pontil. 2004. ``Regularized
multi--task learning.'' In \emph{Proceedings of the 2004 Acm Sigkdd
International Conference on Knowledge Discovery and Data Mining - Kdd
'04}, 109. New York, New York, USA: ACM Press.
\url{https://doi.org/10.1145/1014052.1014067}.

\leavevmode\hypertarget{ref-Fagiolo2007}{}%
Fagiolo, Giorgio, Alessio Moneta, and Paul Windrum. 2007. ``A critical
guide to empirical validation of agent-based models in economics:
Methodologies, procedures, and open problems.'' \emph{Computational
Economics} 30 (3):195--226.
\url{https://doi.org/10.1007/s10614-007-9104-4}.

\leavevmode\hypertarget{ref-Friedman2010}{}%
Friedman, Jerome, Trevor Hastie, and Robert Tibshirani. 2010.
``Regularization Paths for Generalized Linear Models via Coordinate
Descent.'' \emph{Journal of Statistical Software} 33 (1):1--22.
\url{https://doi.org/10.18637/jss.v033.i01}.

\leavevmode\hypertarget{ref-Gourieroux1993}{}%
Gourieroux, C., A. Monfort, and E. Renault. 1993. ``Indirect
Inference.'' \emph{Journal of Applied Econometrics} 8. Wiley:85--118.
\url{https://doi.org/10.2307/2285076}.

\leavevmode\hypertarget{ref-Grazzini2015}{}%
Grazzini, Jakob, and Matteo Richiardi. 2015. ``Estimation of ergodic
agent-based models by simulated minimum distance.'' \emph{Journal of
Economic Dynamics and Control} 51 (February). North-Holland:148--65.
\url{https://doi.org/10.1016/j.jedc.2014.10.006}.

\leavevmode\hypertarget{ref-Grazzini2017}{}%
Grazzini, Jakob, Matteo G. Richiardi, and Mike Tsionas. 2017. ``Bayesian
estimation of agent-based models.'' \emph{Journal of Economic Dynamics
and Control} 77 (April). North-Holland:26--47.
\url{https://doi.org/10.1016/j.jedc.2017.01.014}.

\leavevmode\hypertarget{ref-Grimm2005}{}%
Grimm, Volker, Eloy Revilla, Uta Berger, Florian Jeltsch, Wolf M Mooij,
Steven F Railsback, Hans-Hermann Thulke, Jacob Weiner, Thorsten Wiegand,
and Donald L DeAngelis. 2005. ``Pattern-Oriented Modeling of Agent Based
Complex Systems: Lessons from Ecology.'' \emph{Science} 310 (5750).
American Association for the Advancement of Science:987--91.

\leavevmode\hypertarget{ref-hartig_statistical_2011}{}%
Hartig, F., J.M. Calabrese, B. Reineking, T. Wiegand, and A. Huth. 2011.
``Statistical inference for stochastic simulation models - theory and
application.'' \emph{Ecology Letters} 14 (8):816--27.
\url{https://doi.org/10.1111/j.1461-0248.2011.01640.x}.

\leavevmode\hypertarget{ref-friedman_elements_2001}{}%
Hastie, Trevor, Robert Tibshirani, and Jerome Friedman. 2009. \emph{The
Elements of Statistical Learning}. 1st ed. Vol. 1. Berlin: Springer
series in statistics Springer, Berlin.
\url{https://doi.org/10.1007/b94608}.

\leavevmode\hypertarget{ref-Henningsen2007}{}%
Henningsen, Arne, and Jeff D Hamann. 2007. ``\{systemfit\}: A Package
for Estimating Systems of Simultaneous Equations in \{R\}.''
\emph{Journal of Statistical Software} 23 (4):1--40.
\url{https://doi.org/10.1088/0953-8984/19/36/365219}.

\leavevmode\hypertarget{ref-Heppenstall2007}{}%
Heppenstall, Alison J, Andrew J Evans, and Mark H Birkin. 2007.
``Genetic algorithm optimisation of an agent-based model for simulating
a retail market.'' \emph{Environment and Planning B: Planning and
Design} 34 (6):1051--70. \url{https://doi.org/10.1068/b32068}.

\leavevmode\hypertarget{ref-Jabot2015}{}%
Jabot, Franck, Thierry Faure, Nicolas Dumoulin, and Carlo Albert. 2015.
``EasyABC: Efficient Approximate Bayesian Computation Sampling
Schemes.'' \url{https://cran.r-project.org/package=EasyABC}.

\leavevmode\hypertarget{ref-Kennedy2001a}{}%
Kennedy, Marc C, and Anthony O'Hagan. 2001. ``Bayesian calibration of
computer models.'' \emph{J. R. Stat. Soc. Ser. B Stat. Methodol.} 63
(3):425--64. \url{https://doi.org/10.1111/1467-9868.00294}.

\leavevmode\hypertarget{ref-Klima2018}{}%
Klima, Grzegorz, Karol Podemski, and Kaja Retkiewicz-Wijtiwiak. 2018.
``gEcon: General Equilibrium Economic Modelling Language and Solution
Framework.''

\leavevmode\hypertarget{ref-Lamperti2018}{}%
Lamperti, Francesco, Andrea Roventini, and Amir Sani. 2018.
``Agent-based model calibration using machine learning surrogates.''
\emph{Journal of Economic Dynamics and Control} 90:366--89.
\url{https://doi.org/10.1016/j.jedc.2018.03.011}.

\leavevmode\hypertarget{ref-Le2016}{}%
Le, Vo Phuong Mai, David Meenagh, Patrick Minford, Michael Wickens, and
Yongdeng Xu. 2016. ``Testing Macro Models by Indirect Inference: A
Survey for Users.'' \emph{Open Economies Review} 27 (1). Springer
US:1--38. \url{https://doi.org/10.1007/s11079-015-9377-5}.

\leavevmode\hypertarget{ref-Lee2015}{}%
Lee, Ju Sung, Tatiana Filatova, Arika Ligmann-Zielinska, Behrooz
Hassani-Mahmooei, Forrest Stonedahl, Iris Lorscheid, Alexey Voinov, Gary
Polhill, Zhanli Sun, and Dawn C. Parker. 2015. ``The complexities of
agent-based modeling output analysis.'' \emph{Journal of Artificial
Societies and Social Simulation} 18 (4).
\url{https://doi.org/10.18564/jasss.2897}.

\leavevmode\hypertarget{ref-Marjoram2003}{}%
Marjoram, Paul, John Molitor, Vincent Plagnol, and Simon Tavare. 2003.
``Markov chain Monte Carlo without likelihoods.'' \emph{Proceedings of
the National Academy of Sciences} 100 (26). National Academy of
Sciences:15324--8. \url{https://doi.org/10.1073/pnas.0306899100}.

\leavevmode\hypertarget{ref-Marler2004}{}%
Marler, R T, and J S Arora. 2004. ``Survey of multi-objective
optimization methods for engineering.''
\url{https://doi.org/10.1007/s00158-003-0368-6}.

\leavevmode\hypertarget{ref-McFadden1989}{}%
McFadden, Daniel. 1989. ``A Method of Simulated Moments for Estimation
of Discrete Response Models Without Numerical Integration.''
\emph{Econometrica} 57 (5). Econometric Society:995.
\url{https://doi.org/10.2307/1913621}.

\leavevmode\hypertarget{ref-Michalski2000}{}%
Michalski, Ryszard S. 2000. ``LEARNABLE EVOLUTION MODEL: Evolutionary
Processes Guided by Machine Learning.'' \emph{Machine Learning} 38 (1).
Kluwer Academic Publishers:9--40.
\url{https://doi.org/10.1023/A:1007677805582}.

\leavevmode\hypertarget{ref-Nott2014}{}%
Nott, D. J., Y. Fan, L. Marshall, and S. A. Sisson. 2011. ``Approximate
Bayesian computation and Bayes linear analysis: Towards high-dimensional
ABC.'' \emph{Journal of Computational and Graphical Statistics} 23 (1).
Taylor \& Francis:65--86.
\url{https://doi.org/10.1080/10618600.2012.751874}.

\leavevmode\hypertarget{ref-OHagan2006}{}%
O'Hagan, A. 2006. ``Bayesian analysis of computer code outputs: A
tutorial.'' \emph{Reliability Engineering and System Safety} 91
(10-11):1290--1300. \url{https://doi.org/10.1016/j.ress.2005.11.025}.

\leavevmode\hypertarget{ref-Parry2013}{}%
Parry, Hazel R., Christopher J. Topping, Marc C. Kennedy, Nigel D.
Boatman, and Alistair W.A. Murray. 2013. ``A Bayesian sensitivity
analysis applied to an Agent-based model of bird population response to
landscape change.'' \emph{Environmental Modelling and Software} 45
(July):104--15. \url{https://doi.org/10.1016/j.envsoft.2012.08.006}.

\leavevmode\hypertarget{ref-rashedi_gsa:_2009}{}%
Rashedi, Esmat, Hossein Nezamabadi-pour, and Saeid Saryazdi. 2009.
``GSA: A Gravitational Search Algorithm.'' \emph{Information Sciences},
Special section on high order fuzzy sets, 179 (13):2232--48.
\url{https://doi.org/10.1016/j.ins.2009.03.004}.

\leavevmode\hypertarget{ref-Richiardi2006}{}%
Richiardi, Matteo, Roberto Leombruni, Nicole Saam, and Michele Sonnessa.
2006. ``A Common Protocol for Agent-Based Social Simulation,'' January.
JASSS. \url{http://jasss.soc.surrey.ac.uk/9/1/15.html}.

\leavevmode\hypertarget{ref-Salle2014}{}%
Salle, Isabelle, and Murat Yıldızoğlu. 2014. ``Efficient sampling and
meta-modeling for computational economic models.'' \emph{Computational
Economics} 44 (4). Springer US:507--36.
\url{https://doi.org/10.1007/s10614-013-9406-7}.

\leavevmode\hypertarget{ref-luke_essentials_2009}{}%
Sean, Luke (George Mason University). 2010. \emph{Essentials of
Metaheuristics: A Set of Undergraduate Lecture Notes}. Lulu.
\url{https://doi.org/10.1007/s10710-011-9139-0}.

\leavevmode\hypertarget{ref-shahriari_taking_2016}{}%
Shahriari, Bobak, Kevin Swersky, Ziyu Wang, Ryan P. Adams, and Nando De
Freitas. 2016. ``Taking the human out of the loop: A review of Bayesian
optimization.'' \emph{Proceedings of the IEEE} 104 (1):148--75.
\url{https://doi.org/10.1109/JPROC.2015.2494218}.

\leavevmode\hypertarget{ref-Shalizi2017}{}%
Shalizi, Cosma Rohilla. 2017. ``Indirect Inference.''
\url{http://bactra.org/notebooks/indirect-inference.html}.

\leavevmode\hypertarget{ref-Smith1993}{}%
Smith, A. A. 1993. ``Estimating nonlinear time‐series models using
simulated vector autoregressions.'' \emph{Journal of Applied
Econometrics} 8 (1 S). Wiley Subscription Services, Inc., A Wiley
Company:S63--S84. \url{https://doi.org/10.1002/jae.3950080506}.

\leavevmode\hypertarget{ref-Stow2009}{}%
Stow, Craig A., Jason Jolliff, Dennis J. McGillicuddy, Scott C. Doney,
J. Icarus Allen, Marjorie A.M. Friedrichs, Kenneth A. Rose, and Philip
Wallhead. 2009. ``Skill assessment for coupled biological/physical
models of marine systems.'' \emph{Journal of Marine Systems} 76
(1-2):4--15. \url{https://doi.org/10.1016/j.jmarsys.2008.03.011}.

\leavevmode\hypertarget{ref-Wegmann2009}{}%
Wegmann, Daniel, Christoph Leuenberger, and Laurent Excoffier. 2009.
``Efficient approximate Bayesian computation coupled with Markov chain
Monte Carlo without likelihood.'' \emph{Genetics} 182 (4).
Genetics:1207--18. \url{https://doi.org/10.1534/genetics.109.102509}.

\leavevmode\hypertarget{ref-Zhang2017}{}%
Zhang, Jingjing, Todd E. Dennis, Todd J. Landers, Elizabeth Bell, and
George L.W. Perry. 2017. ``Linking individual-based and statistical
inferential models in movement ecology: A case study with black petrels
(Procellaria parkinsoni).'' \emph{Ecological Modelling} 360:425--36.
\url{https://doi.org/10.1016/j.ecolmodel.2017.07.017}.

\leavevmode\hypertarget{ref-Zhao2010}{}%
Zhao, Linqiao. 2010. ``A Model of Limit-Order Book Dynamics and a
Consistent Estimation Procedure.'' Carnegie Mellon University.

\end{document}